\documentclass[sigconf]{acmart}
\AtBeginDocument{%
  }

\copyrightyear{2026}
\acmYear{2026}
\setcopyright{cc}
\setcctype{by}
\acmConference[ICPP '26]{Proceedings of the 55th International Conference on Parallel Processing}{September 28-October 01, 2026}{Singapore, Singapore}
\acmBooktitle{Proceedings of the 55th International Conference on Parallel Processing (ICPP '26), September 28-October 01, 2026, Singapore, Singapore}
\acmDOI{10.1145/3832810.3832817}
\acmISBN{979-8-4007-2657-6/2026/09}

\usepackage[ruled,vlined]{algorithm2e}
\usepackage{subcaption}

\begin{document}

\title{Hierarchical Shared Memory-Aware Optimization for TRSM on GPU Platforms}

\author{Xinzhe Chen}
\orcid{0009-0000-9295-3623}
\email{chenxinzhe25@mails.ucas.ac.cn}
\affiliation{%
  \institution{Institute of Software, Chinese Academy of Sciences}
  \city{Beijing}
  \country{China}}
\affiliation{%
  \institution{University of Chinese Academy of Sciences}
  \city{Beijing}
  \country{China}}

\author{Haowei Li}
\orcid{0000-0002-1268-6586}
\email{lihaowei22@mails.ucas.ac.cn}
\affiliation{%
  \institution{Institute of Software, Chinese Academy of Sciences}
  \city{Beijing}
  \country{China}}
\affiliation{%
  \institution{University of Chinese Academy of Sciences}
  \city{Beijing}
  \country{China}}

\author{Lijuan Hu}
\orcid{0009-0004-0723-4632}
\email{hulijuan@iscas.ac.cn}
\affiliation{%
  \institution{Institute of Software, Chinese Academy of Sciences}
  \city{Beijing}
  \country{China}}

\author{Wenjing Ma}
\orcid{0000-0002-1795-4498}
\email{wenjing@iscas.ac.cn}
\affiliation{%
  \institution{Institute of Software, Chinese Academy of Sciences}
  \city{Beijing}
  \country{China}}

\author{Fangfang Liu}
\authornote{Corresponding author.}
\orcid{0000-0001-7344-7493}
\email{fangfang@iscas.ac.cn}
\affiliation{%
  \institution{Institute of Software, Chinese Academy of Sciences}
  \city{Beijing}
  \country{China}}

\renewcommand{\shortauthors}{Chen et al.}

\begin{abstract}
Triangular Solve with Multiple Right-hand Sides (TRSM) is a fundamental BLAS Level-3 operation that underpins LU/Cholesky decomposition, sparse direct solvers, and matrix inversion. In the left-side lower-triangular case studied in this paper, efficient GPU implementation remains challenging because forward substitution introduces strict row-wise dependencies, and shared memory is too scarce to hold both operand matrices for wide data types such as double complex. This paper presents HSMA-TRSM, a hierarchical shared memory-aware optimization framework for left-side lower-triangular TRSM on NVIDIA A100, NVIDIA H800, and Hygon DCU Z100 accelerators. For the small-scale regime ($m, n \leq 64$), we design a pipelined compute-memory overlap mechanism through loop unrolling and instruction reordering, and propose a dual thread-group seven-stage pipeline strategy to address shared memory constraints for double complex types. For large-scale problems, we introduce a diagonal block decoupling optimization with an $O(I_B)$ shared-memory footprint for diagonal block inversion, enabling adaptive block size selection based on matrix scale and hardware characteristics. A compile-time configuration selection framework based on offline profiling and online lookup selects the optimal block size per platform with zero runtime overhead. Evaluated on NVIDIA A100, H800, and Hygon DCU Z100, HSMA-TRSM achieves peak speedups of 2.05$\times$ over cuBLAS and 2.06$\times$ over rocBLAS. The gains are strongest in shared-memory-constrained double-complex small cases and in large real-type cases where adaptive blocking improves GEMM-dominated updates, while mature vendor kernels leave less optimization headroom in some regimes.
\end{abstract}




\maketitle

\section{Introduction}
With the widespread deployment of modern GPUs, the Basic Linear Algebra Subprograms (BLAS) library~\cite{lawson1979blas, dongarra1990set} has become critical infrastructure for scientific computing and deep learning applications~\cite{kresse1996efficiency, giannozzi2009quantum, yu2021efficient}. In this work, we study three representative accelerators: NVIDIA A100, NVIDIA H800, and Hygon DCU Z100 (abbreviated as DCU hereafter). Triangular Solve with Multiple Right-hand Sides (TRSM) is a core function in BLAS Level-3, used to solve triangular linear systems of the form $op(A)X = \alpha B$, where $A$ is a triangular matrix and $B$ is a dense matrix containing multiple right-hand sides. In this work, we focus on the left-side lower-triangular case, which is solved by forward substitution and exhibits row-wise dependencies across solution rows. TRSM is widely used in forward/backward substitution of LU and Cholesky decomposition~\cite{anderson1999lapack}, sparse direct solvers~\cite{barrett1994templates}, matrix inversion, and least squares problems, with its execution efficiency directly impacting overall application performance.

Although TRSM is widely used, its efficient GPU implementation faces three classes of challenges. For small-scale matrices ($m, n \leq 64$), vendor implementations load operands into shared memory, but the limited shared-memory/LDS budget per thread block---64KB per block on DCU and 48KB per block on A100/H800 in our target configuration---severely restricts the potential for compute-memory pipelining. In particular, a single $64\times64$ double-complex block already occupies 64KB, leaving no room to stage enough right-hand-side data on DCU and making direct double-buffered designs infeasible on NVIDIA within the per-block budget. For large-scale matrices, vendor libraries fix the block size (rocBLAS fixes it at 128), ignoring matrix scale and hardware heterogeneity; meanwhile, the conventional fused diagonal-block path requires $O(I_B^2)$ shared memory, preventing larger block sizes beyond this limit. Finally, existing optimizations often target a single architecture~\cite{du2012cuda, kondratyuk2021gpu}, limiting performance portability across different GPU platforms.

Addressing the above challenges, this paper proposes HSMA-TRSM, a hierarchical shared memory-aware optimization framework for different GPU platforms, with the following main contributions:
\begin{itemize}
\item A pipelined small-scale TRSM kernel that overlaps data prefetching with computation via loop unrolling and instruction reordering. For double-complex operands, we design a dual thread-group seven-stage pipeline to overcome shared memory constraints and sustain throughput.
\item A diagonal block decoupling strategy for large-scale TRSM that reduces the shared-memory footprint of diagonal block inversion from $O(I_B^2)$ to $O(I_B)$, expanding the feasible block size. Larger block sizes enable the invocation of larger GEMM kernels with higher throughput and reduced kernel launch overhead. An adaptive blocking algorithm then selects the near-optimal block size per matrix scale and data type.
\item A compile-time configuration selection framework based on offline profiling and online lookup, achieving performance portability across NVIDIA A100, H800, and Hygon DCU Z100 with zero runtime overhead. HSMA-TRSM achieves peak speedups of 2.05$\times$ over cuBLAS and 2.06$\times$ over rocBLAS, with performance gains depending on data type, matrix scale, and vendor-library optimization headroom.
\end{itemize}

\section{Background}
\subsection{TRSM Computational Characteristics}
\label{sec:trsm-characteristics}
Triangular linear systems of the form $op(A)X = \alpha B$ arise frequently in numerical linear algebra, where $A$ is an $m \times m$ triangular matrix, $B$ is an $m \times n$ dense matrix, and $op(A)$ denotes an optional transpose operation. Triangular Solve with Multiple Right-hand Sides (TRSM) is the BLAS Level-3 routine~\cite{dongarra1990set} that solves such systems. In this paper, we focus on the left-side lower-triangular case, for which forward substitution computes row by row:
$x_{i,j} = \frac{1}{a_{ii}}\left(\alpha b_{i,j} - \sum_{k=1}^{i-1} a_{ik} x_{k,j}\right)$.
This computation exhibits three key characteristics: solving row $i$ strictly depends on results from the previous $i-1$ rows, forming a serial dependency chain along the solve direction; computational complexity is $O(m^2 n)$ with data volume $O(m^2 + mn)$; each element of triangular matrix $A$ is repeatedly accessed by all $n$ columns of matrix $B$, yielding a data reuse factor of $O(n)$.

\subsection{GPU Architecture and Memory Hierarchy}
Modern GPUs adopt massively parallel SIMT architecture~\cite{hijma2023optimization, zhou2017performance}, hiding memory latency through thousands of lightweight threads. NVIDIA GPUs use Streaming Multiprocessors (SMs) as basic computing units, with threads organized into warps of 32; the Hygon DCU Z100 platform studied in this work uses Compute Units (CUs) as basic units, with threads organized into wavefronts of 64, necessitating platform-specific thread mapping strategies.

The GPU memory hierarchy consists of (from top to bottom): registers, shared memory/Local Data Share (LDS), L1/L2 caches, and global memory (HBM). Shared memory is programmer-managed on-chip storage with substantially lower access latency than global memory, but the effective budget available to a single kernel thread block remains limited in our target setting: 48KB per block on A100/H800 and 64KB per block on DCU.

Modern GPUs support asynchronous memory access instructions (e.g., NVIDIA's `cp.async`), allowing memory operations to overlap with computation in time~\cite{zhang2017understanding}. Fully exploiting this feature enables construction of compute-memory pipelines, prefetching the next data block while computing the current one, thereby hiding memory latency. Furthermore, GPU global memory accesses are coalesced in 128-byte units, and using vectorized data types (e.g., `float4`, `double2`) enables multi-element transfers in a single instruction, significantly improving memory bandwidth utilization~\cite{volkov2008benchmarking}.

\section{Related Work}
\subsection{cuBLAS and rocBLAS Implementations}
The vendor BLAS libraries most relevant here, cuBLAS~\cite{cublas2024} and rocBLAS~\cite{rocblas2024}, both expose hierarchical TRSM implementations. Public rocBLAS source code and our empirical observations of cuBLAS suggest that small-scale problems are kept in shared memory for direct solving, while larger problems are transformed into a sequence of blocked solves and GEMM updates.

For the small-scale regime ($m, n \leq 64$), existing GPU implementations commonly adopt single-thread-block strategies that load sub-blocks of matrices $A$ and $B$ into shared memory for computation. However, these implementations typically separate data loading, computation, and write-back into distinct stages, leaving limited opportunity to overlap memory access with forward-substitution computation in our target left-side lower-triangular setting. For double complex type, where each element occupies 16 bytes, a $64 \times 64$ matrix exactly fills 64KB of shared memory, preventing sufficient caching space for both matrices $A$ and $B$ and leading to frequent global memory accesses.

For large-scale TRSM, mainstream methods adopt recursive blocking strategies, but rocBLAS fixes the block size at 128, with this ``one-size-fits-all'' strategy failing to adaptively adjust based on matrix scale and hardware characteristics. A deeper issue lies in diagonal block solving implementation: rocBLAS employs operator fusion techniques, integrating diagonal block inversion, back-substitution, and small-scale GEMM into a single kernel, which reduces kernel launch overhead but requires $4 \times I_B^2 \times \text{sizeof(datatype)}$ of shared memory space, limiting maximum block size by shared memory capacity. This restriction is particularly detrimental for large-scale problems, as it forces the decomposition into an excessive number of small-granularity GEMM kernels. Since the arithmetic intensity of GEMM scales with the block size, smaller blocks result in lower computational density and fail to effectively amortize memory access costs. Aggregating these operations into larger, more compute-intensive GEMM calls is therefore critical for achieving high hardware utilization and maximizing overall throughput.

Furthermore, vendor BLAS libraries rely heavily on platform-specific hardware features and low-level scheduling strategies, which are often difficult to migrate across different architectures~\cite{charara2017framework, du2012cuda}.

\subsection{Other TRSM Optimization Work}
Prior TRSM work has improved recursive blocking, batching, and portability. Abdelfattah et al.~\cite{abdelfattah2016batched} transformed TRSM/TRMM into GEMM calls through recursive blocking, while Haidar et al.~\cite{haidar2017batched} explored batched triangular solve in MAGMA~\cite{tomov2010magma,gates2014magma} and Dongarra et al.~\cite{dongarra2017batched} proposed a standardized batched BLAS API. Charara et al.~\cite{charara2016redesigning,charara2017framework} redesigned GPU TRMM/TRSM around recursive formulations and KBLAS kernels, and Ringoot et al.~\cite{ringoot2025portable} studied a portable Julia recursive implementation. These works show the value of GEMM-like restructuring and optimized triangular kernels; our focus is narrower and more resource-driven, namely reducing the shared-memory footprint of diagonal-block processing and adapting block sizes across A100, H800, and DCU Z100.

Communication-avoiding TRSM has also been studied from an algorithmic perspective. Wicky et al.~\cite{wicky2017communication} use selective triangular block inversion to reduce communication and synchronization for TRSM with multiple right-hand sides. Their analysis targets parallel communication cost, whereas our implementation targets GPU on-chip resource constraints, especially the $O(I_B^2)$ shared-memory footprint that limits practical block sizes in diagonal-block processing.

Guo et al.~\cite{guo2023satrsm} proposed SA-TRSM for auto-tuning small irregular TRSM shapes, revealing the importance of shape awareness, but it does not address large-scale blocking. More broadly, auto-tuning systems such as CLBlast~\cite{nugteren2018clblast} move tuning cost offline, but most work focuses on compute-intensive GEMM~\cite{whaley1998atlas, li2009auto, tanner2018tensile, su2017automatic, li2022fine}; our offline table is instead used to select TRSM block sizes after diagonal-block decoupling expands the feasible choices.

Existing studies have improved recursive blocking, batched kernels, and auto-tuning for TRSM. However, three issues remain insufficiently addressed for our target setting: shared-memory-constrained pipelining for wide data types in the small-scale regime, the resource cost of diagonal block processing in the large-scale regime, and low-overhead performance-portable tuning across different GPU platforms.

\section{Design}
\subsection{Overall Framework Design}
To address the performance bottlenecks of TRSM on GPUs, this paper proposes a hierarchical shared memory-aware optimization framework (HSMA-TRSM). The core idea is to adaptively select optimal computation strategies and resource allocation schemes based on different problem scales and data types.

We adopt 64 as the unified boundary between the small-scale regime ($m, n \leq 64$) and larger problems. This choice is determined not by shared-memory capacity alone, but by both implementation constraints and the observed performance crossover. Within the small-scale regime, the shared-memory direct-solving strategy avoids the fixed costs of blocked execution, including diagonal-block preprocessing, extra kernel launches, and intermediate data movement. At this scale, the forward-substitution dependency chain, register pressure, and synchronization overhead also remain within an efficient range.

At $128 \times 128$, even for float data, a direct-solving kernel is no longer optimal. The bottleneck is not simply whether the matrix can still be placed on chip, but that the longer dependency chain and the increased on-chip data orchestration and synchronization cost make it difficult for the direct kernel to sustain high efficiency. In contrast, the blocked large-scale algorithm organizes the dominant work into diagonal-block solves and block updates, where the update phase can be transformed into GEMM-like operations with better data reuse and higher arithmetic intensity. Our profiling confirms that, for float data, the blocked path already outperforms the fully unrolled direct-solving kernel at $128 \times 128$.

The $64 \times 64$ double-complex case forms another boundary condition. A conventional small-scale direct-solving kernel cannot be directly applied, because matrix $A$ alone already occupies 64KB of shared memory, leaving insufficient space to simultaneously cache enough sub-blocks of $B$. However, directly switching to the large-scale blocked algorithm is also suboptimal, because the problem contains essentially only one diagonal block, so the preprocessing overhead cannot be effectively amortized by subsequent block updates. Therefore, this case is better handled by a partitioned shared-memory pipeline, which preserves the low-overhead nature of direct solving while relaxing the shared-memory capacity constraint through sub-block partitioning and staged execution.

Therefore, 64 serves as an empirically supported boundary across data types, which naturally partitions TRSM into two optimization regimes:
\begin{itemize}
    \item Small-scale ($m, n \leq 64$): Shared-memory direct solving can be adopted, further subdivided as: fully shared-memory resident ($m \leq 32$) where matrices $A$ and $B$ sub-blocks fully fit in shared memory for all data types; and partially shared-memory resident ($33 \leq m \leq 64$) where only matrix $A$ fits, with Double Complex requiring a special partitioned pipeline strategy.
    \item Large-scale (at least one of $m$ or $n$ exceeds 64): The blocked algorithm becomes more favorable, decomposing the problem into combinations of diagonal-block solves and GEMM updates to better exploit data reuse and arithmetic intensity.
\end{itemize}

\begin{figure}[t]
\centering
\includegraphics[width=0.92\columnwidth]{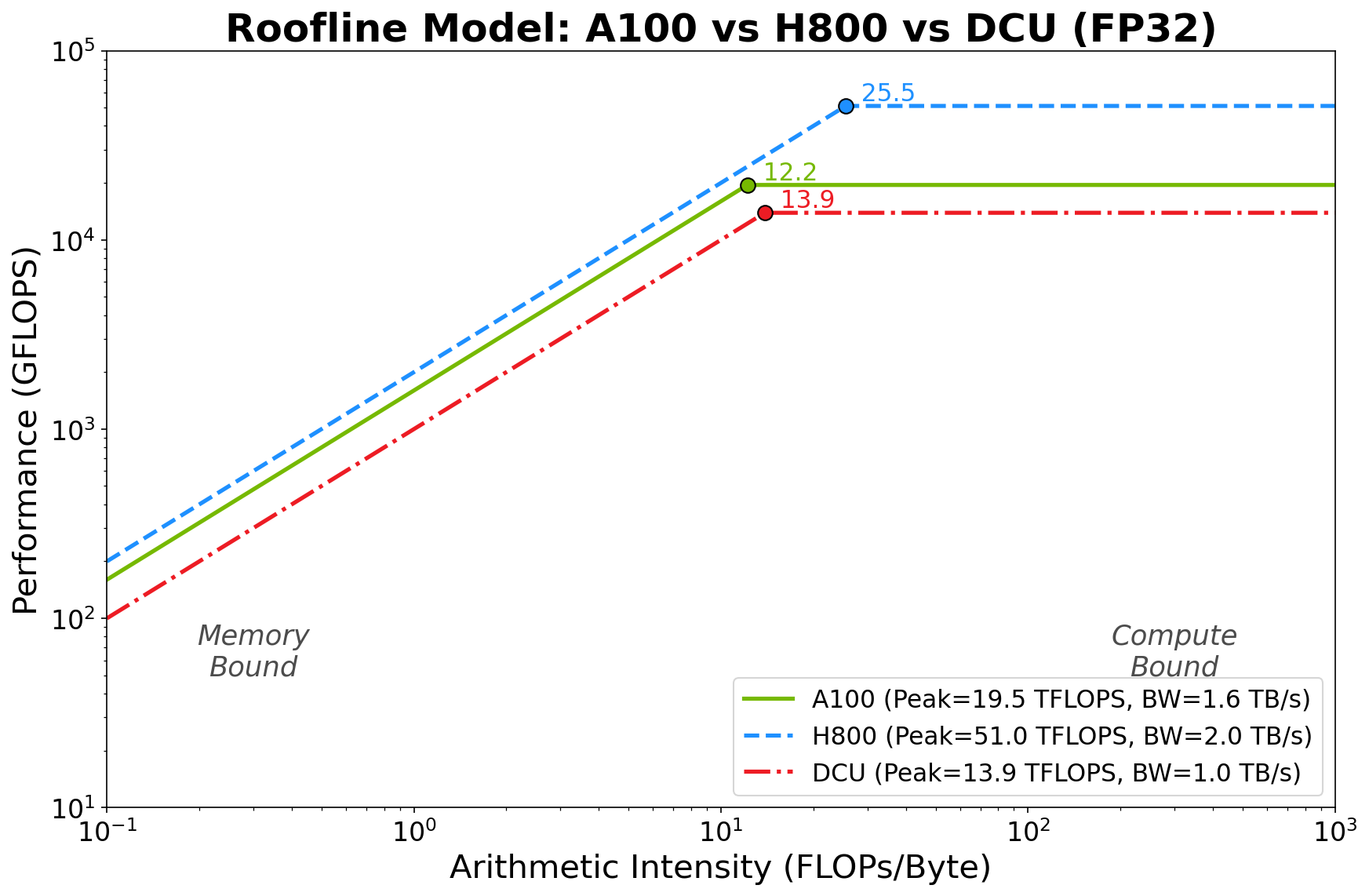}
\caption{Roofline model~\cite{williams2009roofline} analysis of TRSM operations showing memory-bound characteristics for small-scale problems and compute-bound characteristics for large-scale blocked TRSM.}
\Description{A roofline plot with arithmetic intensity on the x-axis and performance (GFLOPS) on the y-axis. Small-scale TRSM operations are positioned in the memory-bound region with low arithmetic intensity, while large-scale blocked TRSM approaches the compute-bound region.}
\label{fig:roofline}
\end{figure}

As shown in Figure~\ref{fig:roofline}, small-scale TRSM exhibits a computational complexity of $O(m^2)$ and a memory traffic of $O(m^2)$, resulting in an arithmetic intensity of approximately $O(1)$---a typical memory-bound computation. Optimization strategies therefore focus on maximizing data reuse, hiding memory latency through compute-memory overlap, and improving memory access efficiency through vectorization. In contrast, large-scale TRSM leverages blocking algorithms to cast the bulk of the computation as GEMM operations~\cite{goto2008high, nath2010improved, tan2011fast, lai2013sgemm}, which possess an arithmetic intensity of $O(n)$ and are thus compute-bound. Optimization strategies focus on selecting appropriate block sizes to maximize GEMM efficiency, optimizing diagonal block solving to reduce serial bottlenecks, and adapting to different problem scales through adaptive tuning.

\begin{figure}[t]
\centering
\includegraphics[width=0.92\columnwidth]{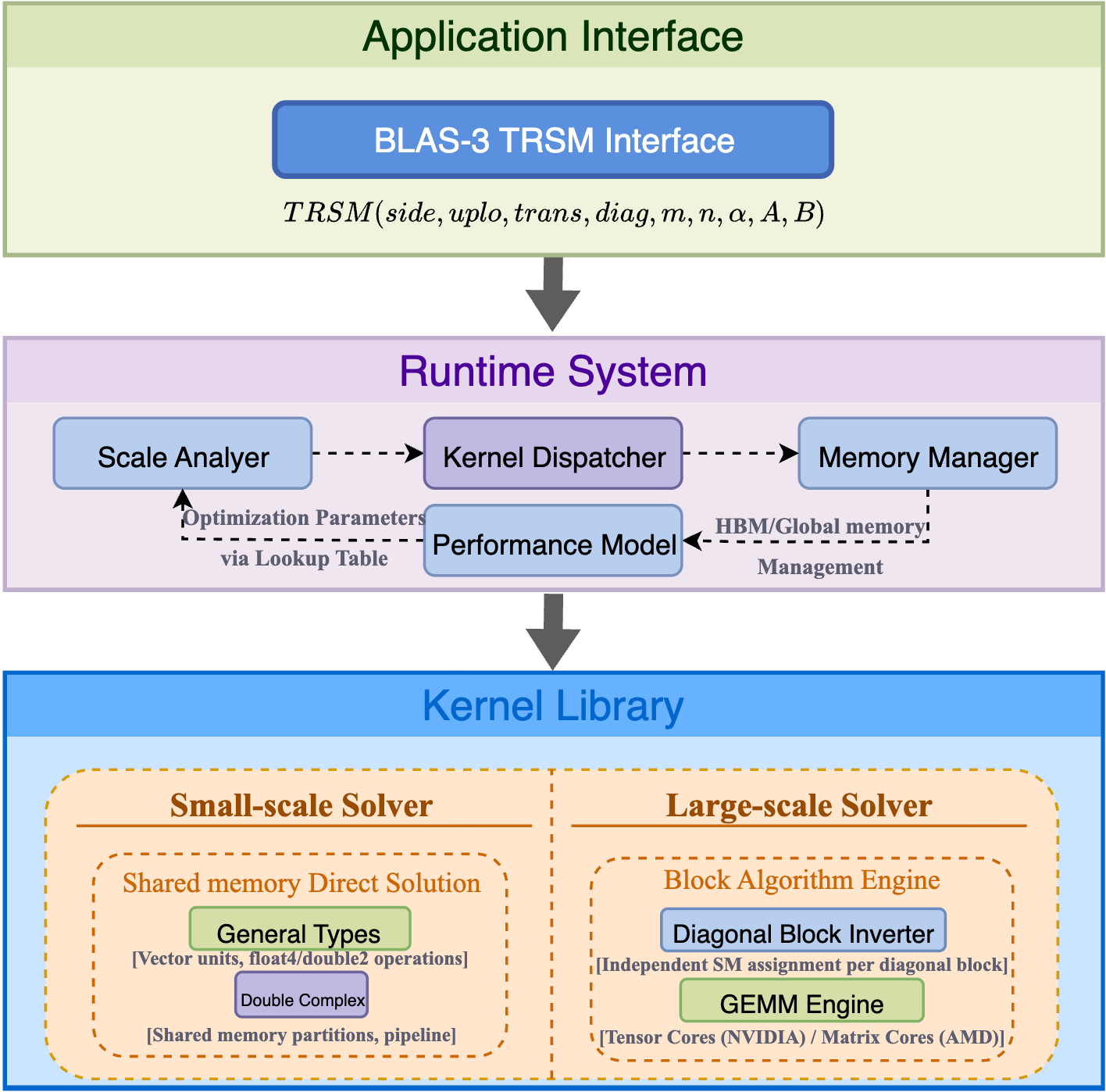}
\caption{Hierarchical optimization framework for TRSM showing adaptive strategy selection based on matrix scale and data type.}
\Description{A flowchart diagram showing the decision tree for TRSM optimization. Input matrix parameters flow through decision nodes that check matrix size and data type, branching to different optimization strategies: shared-memory direct solving for small-scale problems, a partitioned pipeline for double-complex cases within the small-scale regime, and a block-GEMM collaborative strategy for large-scale problems.}
\label{fig:framework}
\end{figure}

The hierarchical optimization framework is shown in Figure~\ref{fig:framework}. The framework automatically selects the optimal computation path based on input matrix scale and data type: small-scale problems use a shared-memory direct solving strategy; within this regime, double-complex cases use a dedicated partitioned pipeline; and large-scale problems use a block-GEMM collaborative strategy.

\subsection{Small-scale TRSM Optimization}
As established in Section~\ref{sec:trsm-characteristics}, the small-scale regime ($m, n \leq 64$) is memory-bound with $O(1)$ arithmetic intensity. The optimizations below therefore target data reuse and memory latency hiding rather than raw computational throughput.

\subsubsection{Shared-Memory Direct Solving Strategy}
\begin{figure}[t]
\centering
\includegraphics[width=0.95\columnwidth]{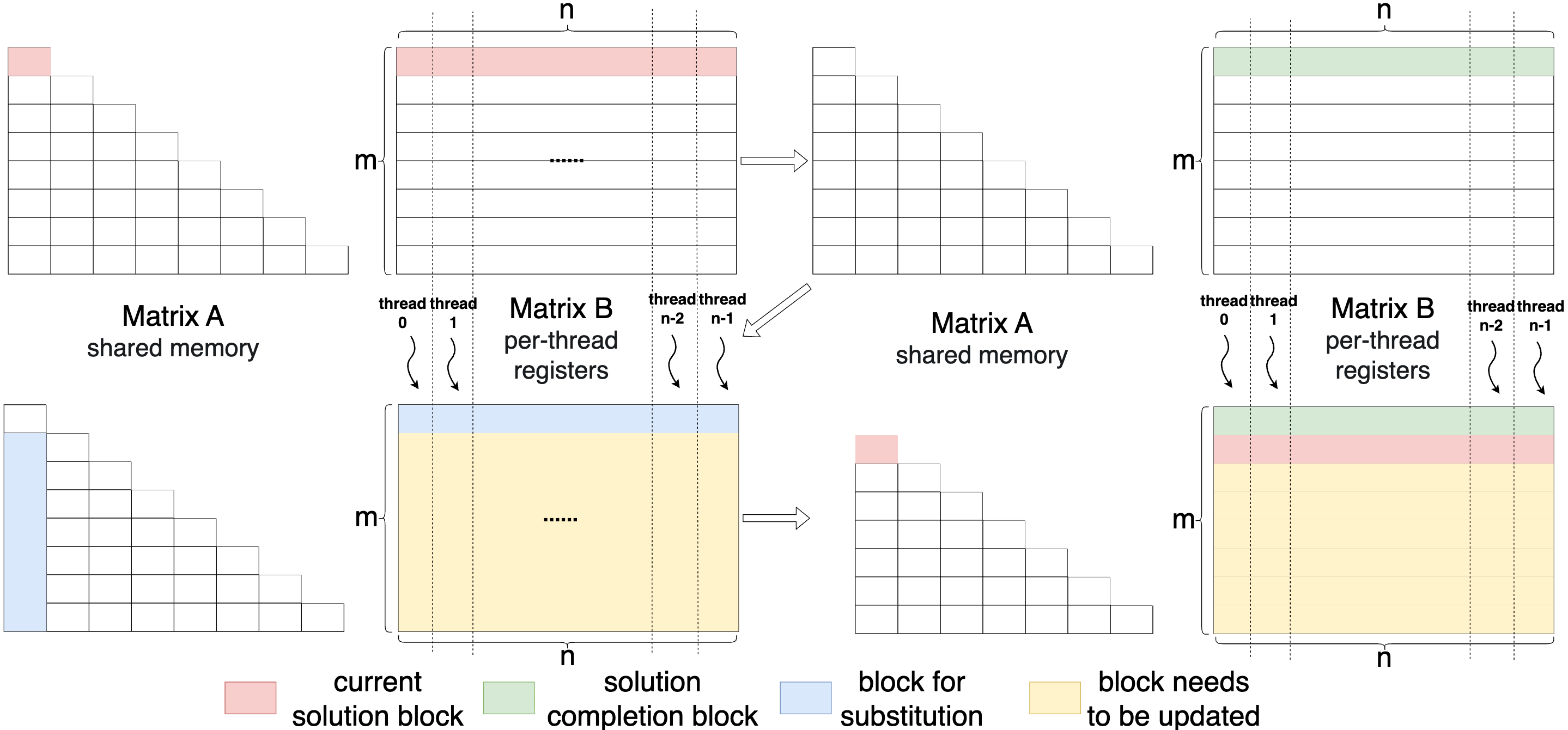}
\caption{Data dependency and column-wise thread mapping in forward substitution. Matrix $A$ is cached in shared memory, while different columns of matrix $B$ are assigned to different threads and kept in registers. Once the current solution row is obtained, it is combined with the corresponding column of matrix $A$ to update the remaining unsolved rows.}
\Description{A diagram illustrating the data dependency and column-wise thread mapping in forward substitution. Matrix A is stored in shared memory, and different columns of matrix B are assigned to different threads and kept in registers. After the current solution row is obtained, it is used together with the corresponding column of matrix A to update the remaining unsolved rows.}
\label{fig:thread-mapping}
\end{figure}

For Float, Double, and Float Complex data types, within the small-scale regime ($m, n \leq 64$), the triangular matrix $A$ can be fully stored in shared memory. In the representative case of $n = 64$ used in Figure~\ref{fig:thread-mapping}, half of the right-hand-side block (32 columns) can also be staged on chip at a time. Based on this organization, we adopt a shared-memory direct solving strategy, achieving high overlap between computation and memory access through multiple optimization techniques.

\paragraph{Forward Substitution Computation Flow }
For the left-side lower-triangular case considered in this paper, forward substitution solves each element of $X$ row by row, with the formula $$x_{i,j} = \frac{1}{a_{ii}}\left(\alpha b_{i,j} - \sum_{k=1}^{i-1} a_{ik} x_{k,j}\right).$$ This process has strict inter-row dependencies: solving row $i$ depends on the results of the previous $i-1$ rows, whereas different columns within the same row are independent and can therefore be processed in parallel. Figure~\ref{fig:thread-mapping} illustrates this data dependency pattern together with the corresponding column-wise thread mapping. Matrix $A$ is cached in shared memory, while different columns of matrix $B$ are assigned to different threads and kept in registers. Once the current solution row is obtained, it is combined with the corresponding column of matrix $A$ to update the remaining unsolved rows. Based on this data organization, vectorized memory instructions (e.g., \texttt{float4} and \texttt{double2}) can be used in the loading stage to improve memory bandwidth utilization.

\paragraph{Compute-Memory Overlap}
Traditional implementations separate data loading, computation, and write-back into coarse-grained phases, preventing memory latency from being hidden by the forward-substitution computation. GPU memory access instructions (e.g., \texttt{ld.shared}) execute asynchronously and stall only at explicit synchronization points. Based on this characteristic, after matrix $A$ is cached in shared memory, we restructure the row-wise solving flow of matrix $B$ so that loading $B[i+1]$ can overlap with solving row $i$, while write-back operations are delayed until the corresponding rows are no longer needed by subsequent updates.

\begin{figure}[t]
\centering
\includegraphics[width=0.95\columnwidth]{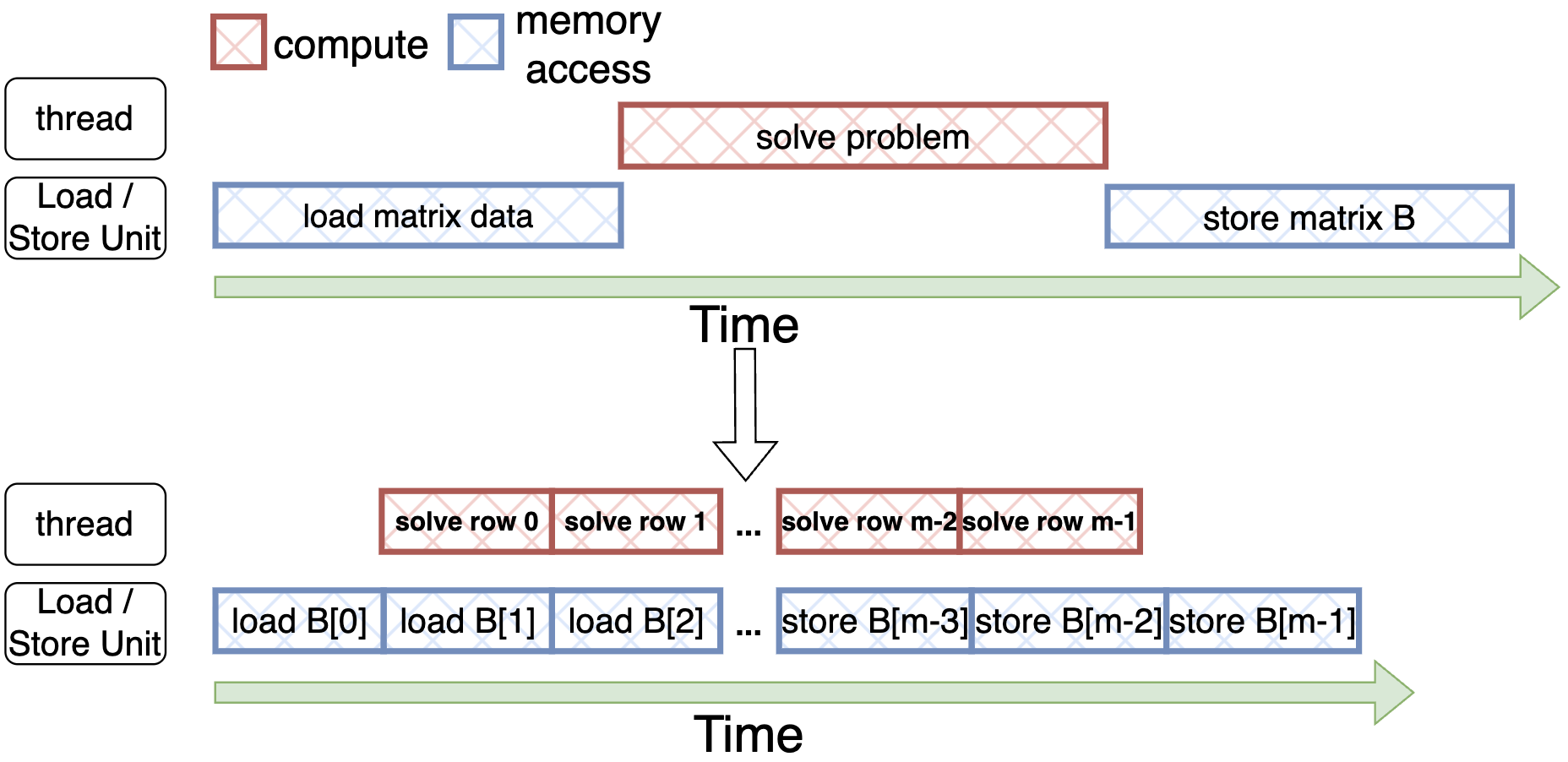}
\caption{Timeline comparison between the traditional coarse-grained execution and the optimized row-wise pipeline. The optimized implementation overlaps loading $B[i+1]$ with solving row $i$ and delays write-back operations to form a deeper load-compute-store pipeline.}
\Description{A timeline diagram comparing traditional and optimized implementations. The traditional timeline separates matrix loading, solving, and storing into coarse phases. The optimized timeline decomposes the process into row-wise operations, overlapping the load of the next row of matrix B with the solve of the current row and delaying stores until the corresponding rows are no longer needed.}
\label{fig:pipeline-timing}
\end{figure}

However, a simple overlap between adjacent iterations is still insufficient: although part of the computation within a single iteration can be covered, the next iteration still waits for its required data to arrive, leaving pipeline bubbles between adjacent rows. To further reduce these bubbles, we adopt a complete loop unrolling strategy~\cite{rocha2020vectorization} and reorder instructions by moving future row loads earlier and deferring completed row write-backs. This forms a deeper pipeline across load, compute, and write-back stages while maintaining constant communication volume per iteration. Figure~\ref{fig:pipeline-timing} illustrates the timeline comparison between the traditional coarse-grained execution and the optimized row-wise pipeline. Algorithm~\ref{alg:solve} details the solving flow after loop unrolling and instruction reordering.

\begin{algorithm}[t]
\caption{Solving Algorithm with Loop Unrolling and Instruction Reordering}
\label{alg:solve}
\KwIn{Matrix $A$, Matrix $B$}
\KwOut{Solution matrix $B$}
Load $A$ to shared memory $sA$\;
\tcp{loop 0}
Load $B[0]$ to register $resB[0]$\;
Solve row 0\;
Load $B[1]$ to register $resB[1]$\;
\tcp{loop 1}
Load $B[2]$ to register $resB[2]$\;
Solve row 1\;
Load $B[3]$ to register $resB[3]$\;
$\cdots$\;
\tcp{loop m-1}
Write $resB[m-2]$ back to global memory $B[m-2]$\;
Solve row $m-1$\;
Write $resB[m-1]$ back to global memory $B[m-1]$\;
\end{algorithm}

\subsubsection{Partitioned Pipeline Strategy}
Double Complex type occupies 16 bytes per element, and a $64 \times 64$ matrix exactly fills 64KB of shared memory ($64^2 \times 16 = 65536$ bytes), meaning shared memory can only accommodate either matrix $A$ or $B$, not both simultaneously. We therefore propose a dual thread-group pipeline strategy based on shared-memory partitioning.

This design is a boundary-case compromise rather than a generally preferred strategy. At $m=64$, a full on-chip direct solve is no longer feasible, while directly switching to the large-scale blocked path would introduce diagonal-block setup, kernel-launch, and GEMM overhead that cannot yet be amortized. We therefore preserve on-chip reuse for the critical triangular data while streaming the remaining operands in smaller pieces. This rationale does not extend to larger float or double problems, where the blocked formulation exposes substantially higher throughput by turning most of the work into larger GEMM updates.

Specifically, we partition the 64KB shared memory into four 16KB sub-regions for storing sub-blocks of matrix $A$ ($A_{11}$, $A_{21}$, $A_{22}$) and matrix $B$ ($B_1$, $B_2$, $B_3$, $B_4$). Matrix $A$ is partitioned in a $2\times 2$ block pattern, and matrix $B$ in a $4 \times 1$ block pattern, allowing each sub-block to fit in the partitioned shared memory.

\begin{figure*}[t]
\centering
\begin{subfigure}[b]{0.24\textwidth}
  \centering
  \includegraphics[width=\textwidth]{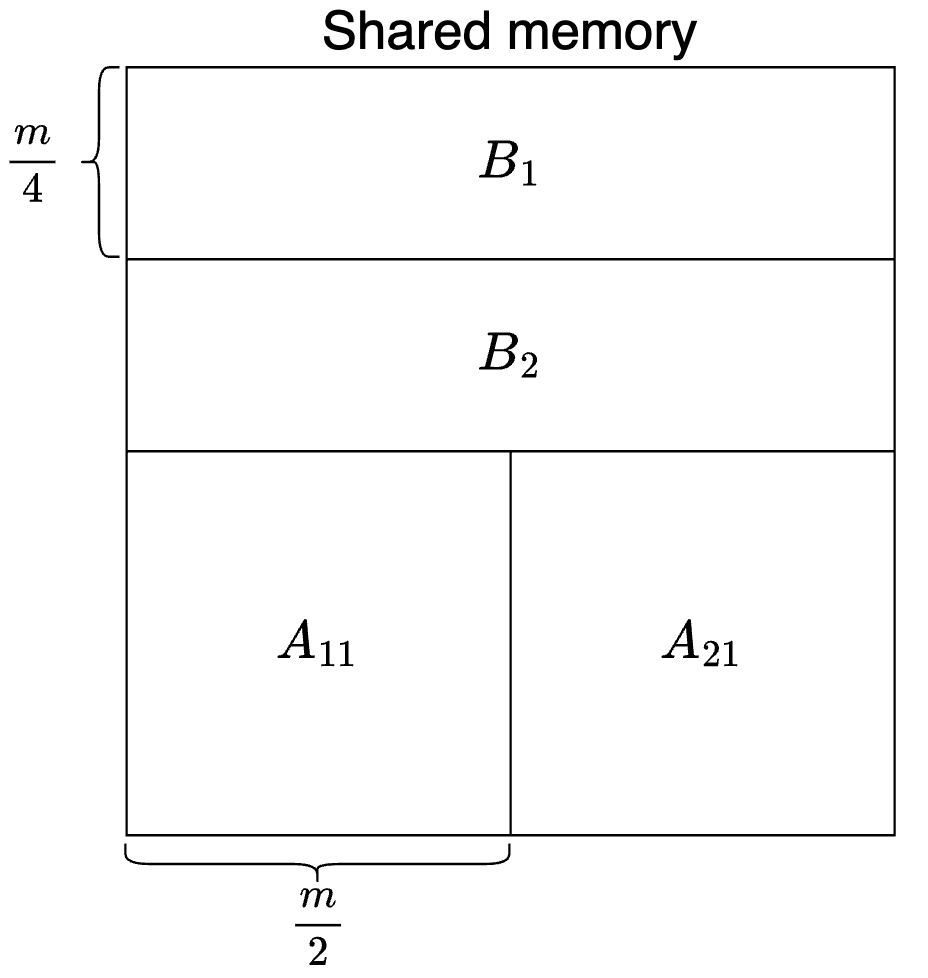}
  \caption{Shared memory partitioning scheme}
  \label{fig:memory-partition}
\end{subfigure}
\hspace{0.01\textwidth}
\begin{subfigure}[b]{0.73\textwidth}
  \centering
  \includegraphics[width=\textwidth]{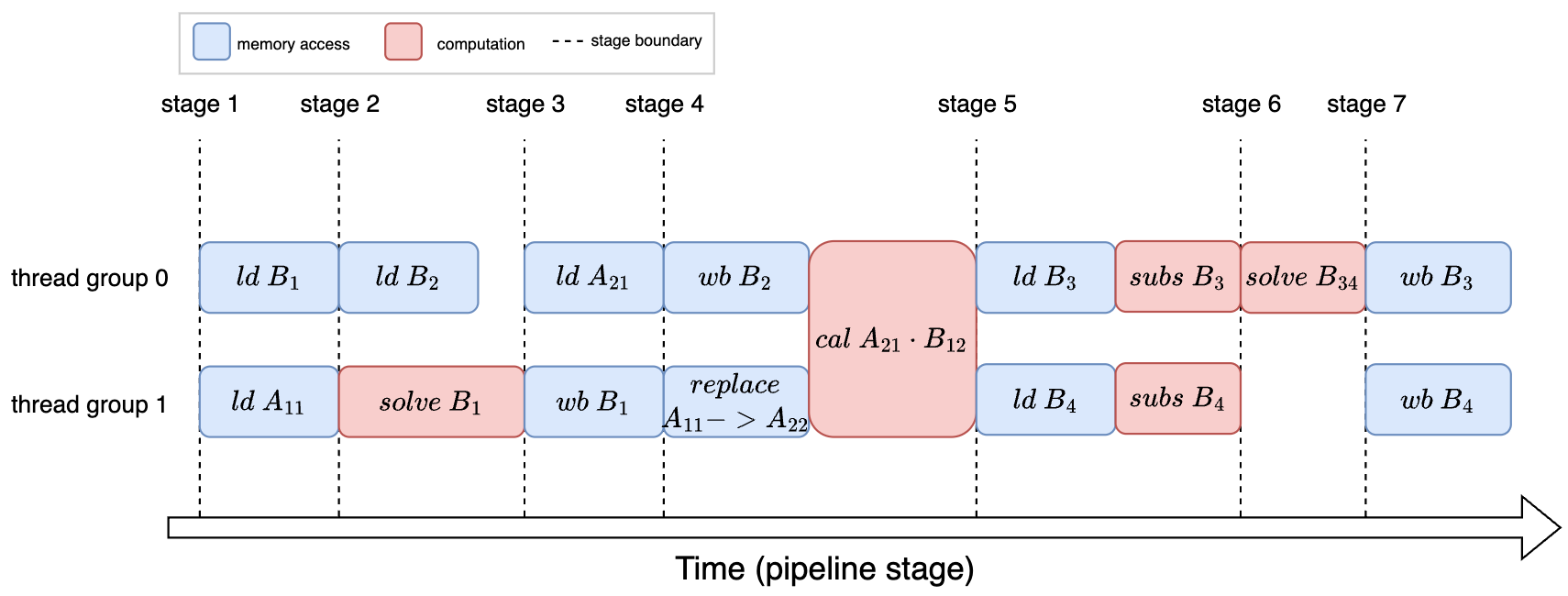}
  \caption{Seven-stage dual thread-group pipeline}
  \label{fig:dual-pipeline}
\end{subfigure}
\caption{Double complex TRSM optimization: (a) shared memory partitioning; (b) seven-stage dual thread-group pipeline.}
\Description{Left: shared memory divided into four 16KB regions for matrix sub-blocks. Right: pipeline diagram showing seven stages with two thread groups overlapping memory and compute operations.}
\label{fig:double-complex-opt}
\end{figure*}

We employ 128 threads divided into two thread groups working collaboratively, constructing a seven-stage pipeline. The first stages load $A_{11}$ together with $B_1$ and $B_2$, then solve and write back $B_1$. The middle stages stream $A_{21}$ and $A_{22}$, use $A_{21} \times B_{1,2}$ to update the remaining sub-blocks, and load $B_3$ and $B_4$. The final stages solve the updated $B_3$ and $B_4$ with $A_{22}$ and write them back. Figure~\ref{fig:double-complex-opt} illustrates both the shared memory partitioning scheme and the seven-stage pipeline.

By assigning memory operations to one thread group and computation to the other, the pipeline achieves full compute-memory overlap despite the shared memory capacity constraint.

\subsection{Large-scale TRSM Optimization}
Large-scale TRSM employs blocking algorithms to decompose the original problem into two execution phases~\cite{tomov2010dense, charara2017framework}.

\paragraph{Phase 1 Parallel Diagonal Block Inversion}
The triangular matrix $A$ is partitioned along the diagonal into $k+1$ diagonal blocks of size $N_B \times N_B$: $D_0, D_1, \ldots, D_k$. Since diagonal blocks are mutually independent, these $k+1$ tasks can be launched simultaneously to compute their inverse matrices $D_0^{-1}, D_1^{-1}, \ldots, D_k^{-1}$ in parallel.

\paragraph{Phase 2 Sequential GEMM Update and Solve}
For the $r$-th block ($r = 0, 1, \ldots, k$), the solving process is:
\begin{itemize}
    \item[1.] GEMM Update (when $r > 0$): $B_r \leftarrow B_r - A_{r,[0:r]} \times X_{[0:r]}$, where $A_{r,[0:r]}$ represents all off-diagonal blocks to the left of the $r$-th diagonal block, and $X_{[0:r]}$ represents the first $r$ solved result blocks. Note that the $K$ dimension of GEMM is $r \times N_B$, growing linearly with iteration count.
    \item[2.] Diagonal Block Solve: $X_r = D_r^{-1} \times B_r$, completed using the inverse matrix pre-computed in Phase 1.
\end{itemize}

Since solving $X_r$ depends on $X_0, X_1, \ldots, X_{r-1}$, Phase 2 must execute sequentially. This two-phase design removes the diagonal block inversion stage from the serial path, leaving GEMM updates and diagonal block solves in the sequential phase. However, the solve phase after inversion remains a bottleneck in forward substitution, and the efficiency of diagonal block processing still directly affects overall performance through its impact on block size and precomputation cost.

Off-diagonal block GEMM updates can directly call highly optimized GEMM kernels. This section first analyzes resource constraint issues in diagonal block processing and proposes double buffering-based optimization strategies; then introduces adaptive blocking algorithms that dynamically select optimal block sizes based on matrix scale and data type.

\subsubsection{Diagonal Block Optimization}
In blocked TRSM algorithms, block size selection has a decisive impact on overall performance. Larger block sizes can improve GEMM computational efficiency and reduce kernel launch overhead; however, block size is constrained by shared memory consumption during diagonal block inversion.

\begin{figure}[t]
\centering
\includegraphics[width=0.92\columnwidth]{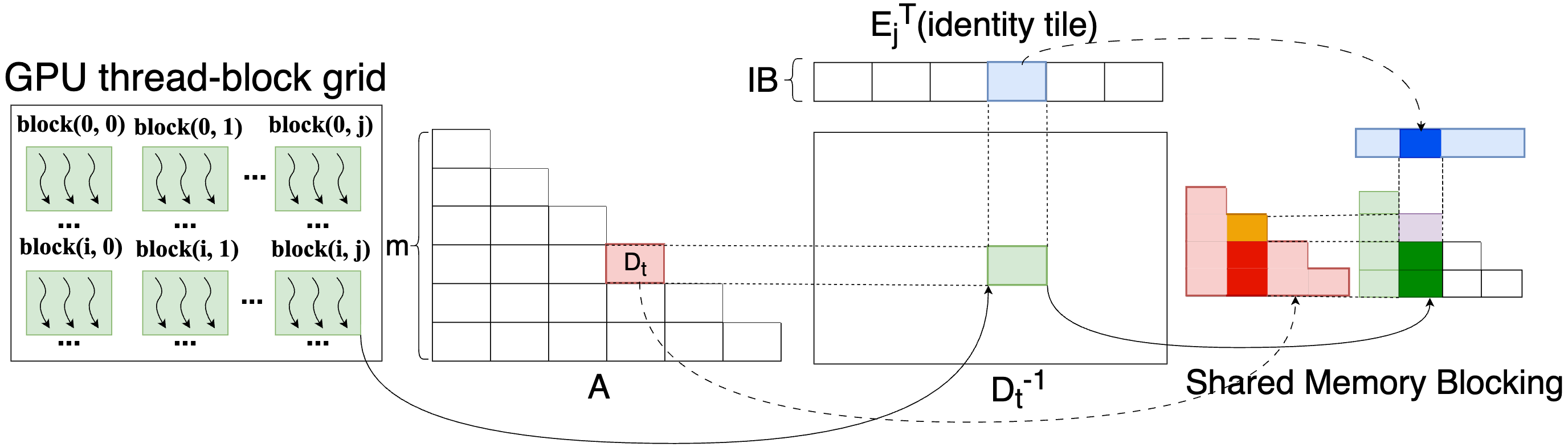}
\caption{Parallel diagonal block inversion and GPU mapping for large-scale TRSM. Independent diagonal blocks $D_t$ are assigned to GPU thread blocks for parallel inversion. Within one diagonal-block task, $E_j^T$ denotes an identity tile used as the right-hand side to generate the corresponding columns or sub-blocks of $D_t^{-1}$, while the shared-memory blocking layout keeps only the currently required data for column-level processing.}
\Description{A diagram illustrating parallel diagonal block inversion and its GPU mapping. One highlighted diagonal block Dt of matrix A is selected for inversion, and GPU thread blocks compute the corresponding sub-blocks of Dt inverse. The label Ej transpose denotes an identity tile used as the right-hand side for generating columns or sub-blocks of the inverse. The shared-memory blocking layout keeps only the currently required triangular sub-block, identity tile, and intermediate results to support column-level processing with reduced memory usage.}
\label{fig:diagonal-decoupling}
\end{figure}

The diagonal block inversion process itself also employs a blocking strategy. Let the outer block size be $N_B$; each diagonal block $D_t$ is further partitioned into $I_B \times I_B$ sub-blocks for recursive solving, where $I_B$ denotes the inner block size, typically set to $N_B/8$. Because different diagonal blocks are independent, they can be assigned to different GPU thread blocks and inverted in parallel. Figure~\ref{fig:diagonal-decoupling} illustrates this mapping: the highlighted block $D_t$ denotes one diagonal block to be inverted, while block$(i,j)$ in the GPU thread-block grid computes the corresponding sub-block of $D_t^{-1}$. The label $E_j^T$ denotes an identity tile used as the right-hand side for generating the corresponding columns or sub-blocks of $D_t^{-1}$. This two-level blocking structure allows diagonal block inversion to reuse small-scale TRSM optimization implementations while transforming inter-sub-block updates into matrix multiplication operations through blocked matrix inversion formulas.

Traditional implementations employ operator fusion techniques, integrating $2 \times I_B$ scale diagonal block inversion, back-substitution, and small-scale GEMM into a single kernel. While this reduces kernel launch overhead, it requires $4 \times I_B^2 \times \text{sizeof(datatype)}$ of shared memory space. Taking Double Complex type as an example, when $I_B = 64$, diagonal block processing alone requires $4 \times 64^2 \times 16 = 256$ KB of shared memory. Following the unified 64KB per-block DCU reference used throughout this paper, this requirement already exceeds the available on-chip budget by a wide margin; under the 48KB per-block budget used by our NVIDIA kernels, the constraint is even tighter. As a result, the maximum feasible block size is strictly constrained to a small range.

To break through this limitation, we decouple the diagonal block computation flow, reducing the shared-memory complexity from $O(I_B^2)$ to $O(I_B)$. Specifically, we remove the inefficient small-scale GEMM module and restructure the computation as a pure back-substitution-based diagonal block inversion algorithm. Since back-substitution has strict inter-column dependencies, each step only requires the current column and the previously solved columns, without loading the entire diagonal block into shared memory simultaneously. Based on this characteristic, we refine the data-loading granularity to the column level and keep only the currently required triangular sub-block, identity tile, and intermediate results in shared memory, as shown on the right side of Figure~\ref{fig:diagonal-decoupling}. In this way, only two adjacent columns need to be loaded at a time, reducing the shared-memory requirement to $2 \times I_B \times \text{sizeof(datatype)}$.

\begin{figure}[t]
\centering
\includegraphics[width=0.50\columnwidth]{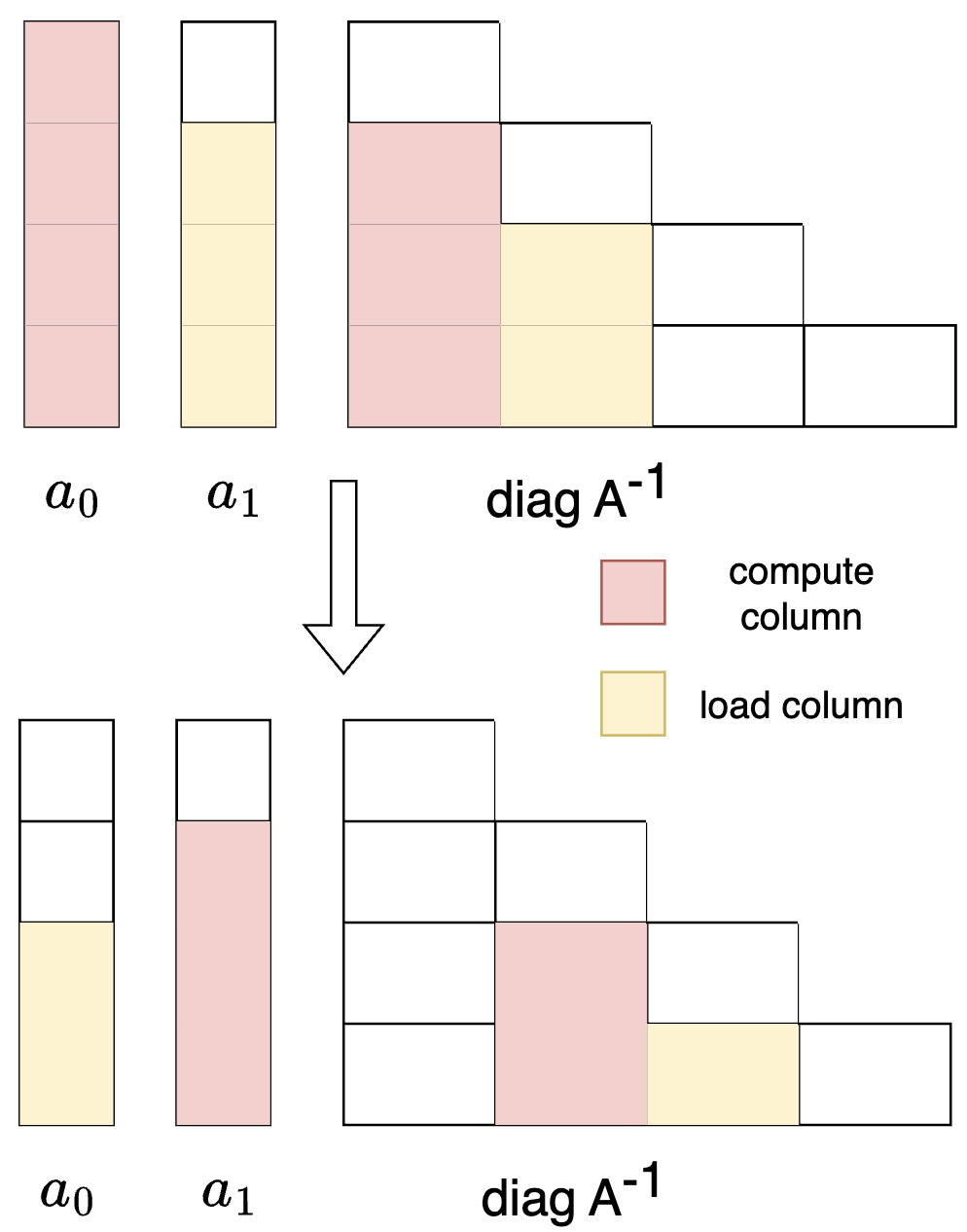}
\caption{Two-step alternation of the double-buffering mechanism for diagonal block inversion. Buffers $a_0$ and $a_1$ alternate between computing the current column of the diagonal-block inverse and prefetching the next column, thereby overlapping data loading with computation.}
\Description{A two-step diagram illustrating double buffering for diagonal block inversion. In the first step, buffer a0 is used to compute the current column of the diagonal-block inverse while buffer a1 loads the next column. In the second step, the roles of the two buffers are exchanged. This alternation overlaps data loading with computation as the columns of the inverse are generated sequentially.}
\label{fig:double-buffer}
\end{figure}

However, a pure column-level loading strategy leads to decreased compute unit utilization, as each computation---encompassing both the update of remaining elements using previously solved values and the solving of the current unknown---must wait for the current column's memory access to complete. To address this, we introduce a Double Buffering mechanism to construct an asynchronous computation pipeline (Figure~\ref{fig:double-buffer}). As shown in Algorithm~\ref{alg:double-buffer}, shared memory is divided into two buffers, denoted as $a_0$ and $a_1$ in the figure. In one step, one buffer is used to compute the current column of the diagonal-block inverse while the other buffer prefetches the next column; in the next step, their roles are exchanged. Through this two-step alternation, data loading and computation operations overlap in time, effectively hiding memory access latency.

\begin{algorithm}[t]
\caption{Double Buffering Diagonal Block Inversion Algorithm}
\label{alg:double-buffer}
\KwIn{Diagonal block matrix $A$}
\KwOut{Inverse matrix $A^{-1}$}
\texttt{\_\_shared\_\_ T sA[2][2*I\_B]}\;
\tcp{prefetch first column}
Load $A[:,0]$ into $sA[0]$\;
\For{$col \leftarrow 0$ \KwTo $2 \times I_B - 1$}{
    Load $A[:,col+1]$ into $sA[(col+1) \mod 2]$ \tcp*{async prefetch}
    Solve column $col$ using $sA[col \mod 2]$\;
    \_\_syncthreads()\;
}
\end{algorithm}

Experimental verification shows that while this optimization strategy increases shared memory requirements from $2 \times I_B$ to $4 \times I_B$ (for double buffering), it still maintains $O(I_B)$ space complexity, saving substantial storage overhead compared to the original $O(I_B^2)$ scheme. This dramatic reduction in shared memory usage enables significantly larger block sizes, providing ample optimization space for subsequent adaptive blocking.

\subsubsection{Adaptive Blocking Algorithm }
\begin{figure*}[t]
\centering
\includegraphics[width=0.84\textwidth]{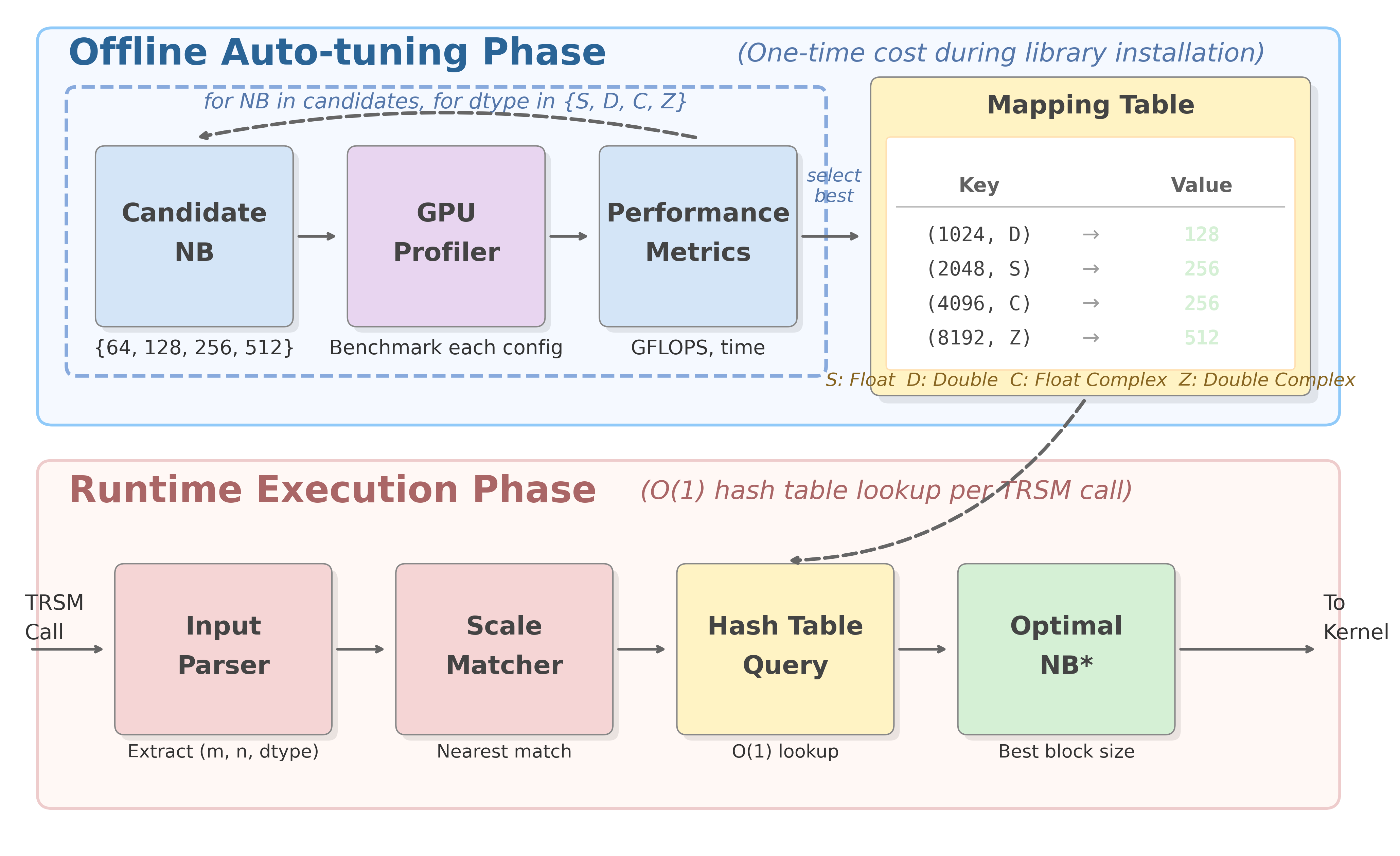}
\caption{Adaptive blocking workflow with offline profiling and online lookup.}
\Description{A flowchart showing the two-phase adaptive blocking process. The offline phase shows traversal of candidate block sizes with performance profiling to build an optimal configuration table. The online phase shows runtime lookup based on input matrix dimensions and data type to select the best block size.}
\label{fig:adaptive-blocking}
\end{figure*}

Since the diagonal blocking optimization alleviated the constraint on block size, we further propose an adaptive blocking algorithm that dynamically selects optimal block sizes based on input matrix and data type, as shown in Figure~\ref{fig:adaptive-blocking}. Block size selection involves multiple trade-offs~\cite{li2019coordinated}: blocks that are too small increase kernel launch counts and overhead; blocks that are too large, while improving single GEMM efficiency, may cause load imbalance or exceed hardware resource limits.

We adopt a two-phase strategy combining offline profiling with online lookup~\cite{nugteren2018clblast, wei2022iatf}. In the pre-build phase, for target hardware platforms, we traverse candidate block size sets (e.g., 32, 64, 128, 256, etc.), execute benchmark tests for each block size, and record actual performance under different matrix scale and data type combinations. Test results are filtered to retain only the optimal block sizes for each configuration, constructing a "scale $\times$ data type $\rightarrow$ optimal block" mapping table.

At runtime, the blocking controller first parses the input matrix dimensions and data type, then looks up the closest scale configuration in the mapping table and returns the corresponding optimal block size. Since the pre-build phase can only cover scale points, for scales not present in the mapping table, the controller adopts a nearest-match strategy, selecting the block size corresponding to the tested scale closest to the actual input scale.

The advantage of this strategy is that it completely transfers performance tuning computational overhead to the offline phase, requiring only a hash table lookup at runtime without introducing additional performance overhead. Compared to rocBLAS's fixed block size of 128, adaptive blocking can select more appropriate configurations for different large-matrix scenarios, avoiding unnecessary blocked-update overhead near the crossover region while using larger blocks to improve GEMM efficiency once the matrix size grows.

\section{Performance Evaluation}
This section evaluates the performance of the proposed TRSM optimization strategies. The experimental platforms include three GPU accelerators: NVIDIA A100-SXM4-40GB, NVIDIA H800 PCIe, and Hygon DCU Z100~\cite{zhou2023dcu}, representing hardware environments with different architectures and computational capabilities; detailed specifications are listed in Table~\ref{tab:hardware}. The comparison baselines are highly optimized BLAS libraries provided by vendors: cuBLAS v13.2.1~\cite{cublas2024} on NVIDIA platforms and rocBLAS v5.1~\cite{rocblas2024} on the DCU platform.

\subsection{Experimental Setup}
\begin{table*}[t]
\centering
\caption{Hardware Platform Specifications}
\label{tab:hardware}
\begin{tabular}{lcccccc}
\toprule
Platform & SM/CU & Shmem/LDS per Block & HBM & Bandwidth & FP32 Peak & Thread Group \\
\midrule
A100 & 108 SM & 48KB & 40GB HBM2e & 1.6 TB/s & 19.5 TFLOPS & 32 (warp) \\
H800 & 132 SM & 48KB & 80GB HBM3 & 2.0 TB/s & 51 TFLOPS & 32 (warp) \\
Hygon DCU Z100 & 64 CU & 64KB & 16GB HBM2 & 1 TB/s & 13.9 TFLOPS & 64 (wavefront) \\
\bottomrule
\end{tabular}
\end{table*}

\paragraph{Test Configuration:} Performance is evaluated on square matrices ($m = n$), with input matrices initialized using random values. We report results for four data types (float, double, float complex, and double complex) in both the small-scale and large-scale regimes. Each test point runs 10 times, and we report the average of the last 9 runs, using the first run as warmup.

\begin{table}[t]
\centering
\caption{Normwise relative output error against cuBLAS on A100}
\label{tab:numerical-error}
\scriptsize
\resizebox{\columnwidth}{!}{%
\begin{tabular}{lccc}
\toprule
Data type & $m=n=1024$ & $m=n=4096$ & $m=n=8192$ \\
\midrule
Float & $1.47{\times}10^{-7}$ & $3.54{\times}10^{-7}$ & $5.14{\times}10^{-7}$ \\
Double & $2.42{\times}10^{-16}$ & $6.83{\times}10^{-16}$ & $9.82{\times}10^{-16}$ \\
Float complex & $1.45{\times}10^{-7}$ & $3.77{\times}10^{-7}$ & $5.74{\times}10^{-7}$ \\
Double complex & $4.03{\times}10^{-16}$ & $8.96{\times}10^{-16}$ & $1.08{\times}10^{-15}$ \\
\bottomrule
\end{tabular}}
\end{table}

\paragraph{Numerical Correctness:} To validate the diagonal-inversion-based path, we compare the output with cuBLAS on A100 using $\|X_{\mathrm{ours}}-X_{\mathrm{cuBLAS}}\|_F/\|X_{\mathrm{cuBLAS}}\|_F$. Table~\ref{tab:numerical-error} shows errors on the order of $10^{-7}$ for single-precision types and $10^{-16}$--$10^{-15}$ for double-precision types, indicating numerical consistency with cuBLAS on the tested cases.

\subsection{Small-Scale TRSM Performance}
The small-scale regime ($m, n \leq 64$) adopts a shared-memory direct solving strategy, improving performance through optimization techniques such as loop unrolling, instruction reordering, and compute-memory overlap.

\subsubsection{Performance Analysis}

\begin{figure}[t]
\centering
\begin{subfigure}[b]{0.46\columnwidth}
  \includegraphics[width=\textwidth]{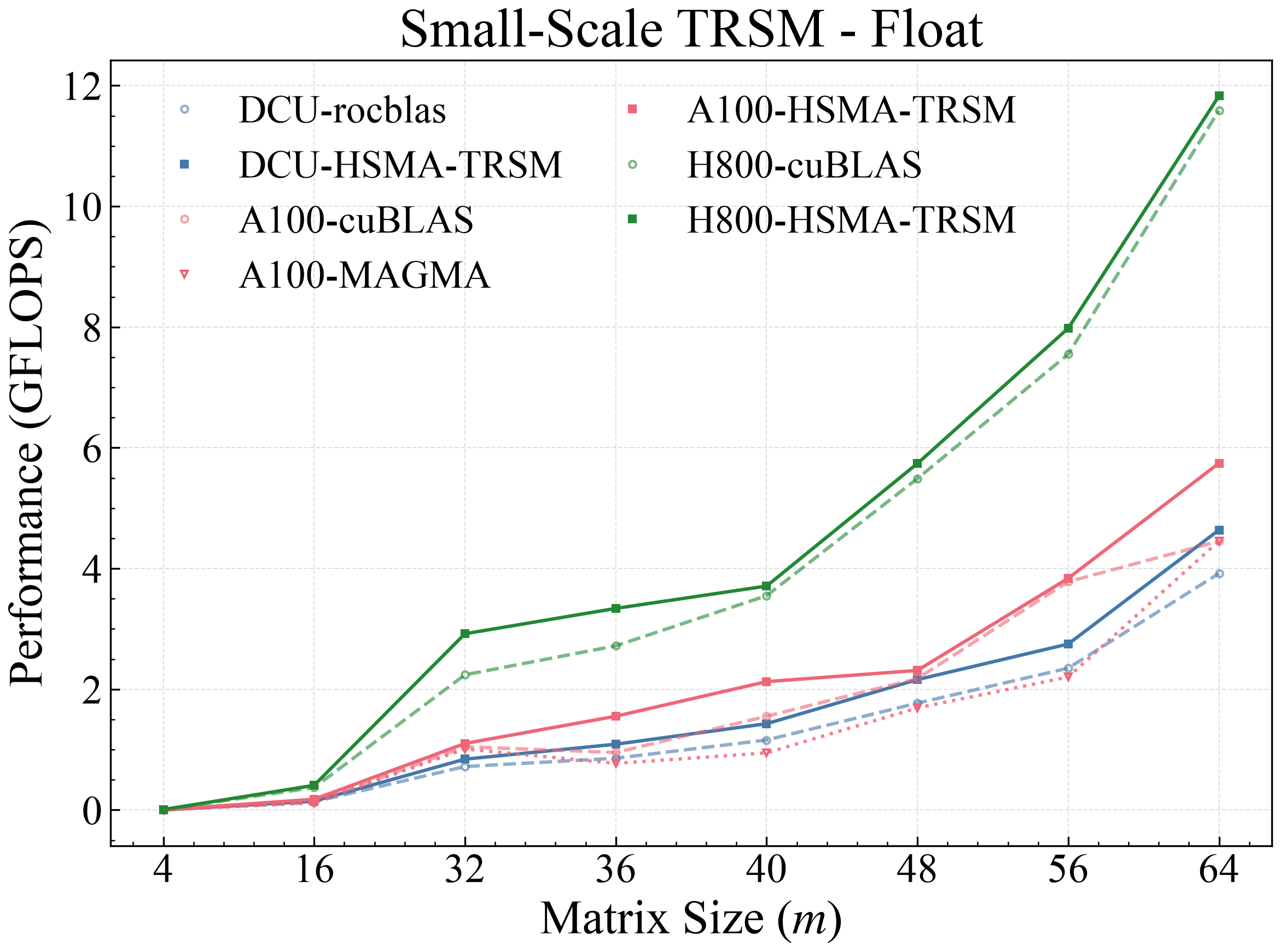}
  \caption{Float}
\end{subfigure}
\hfill
\begin{subfigure}[b]{0.46\columnwidth}
  \includegraphics[width=\textwidth]{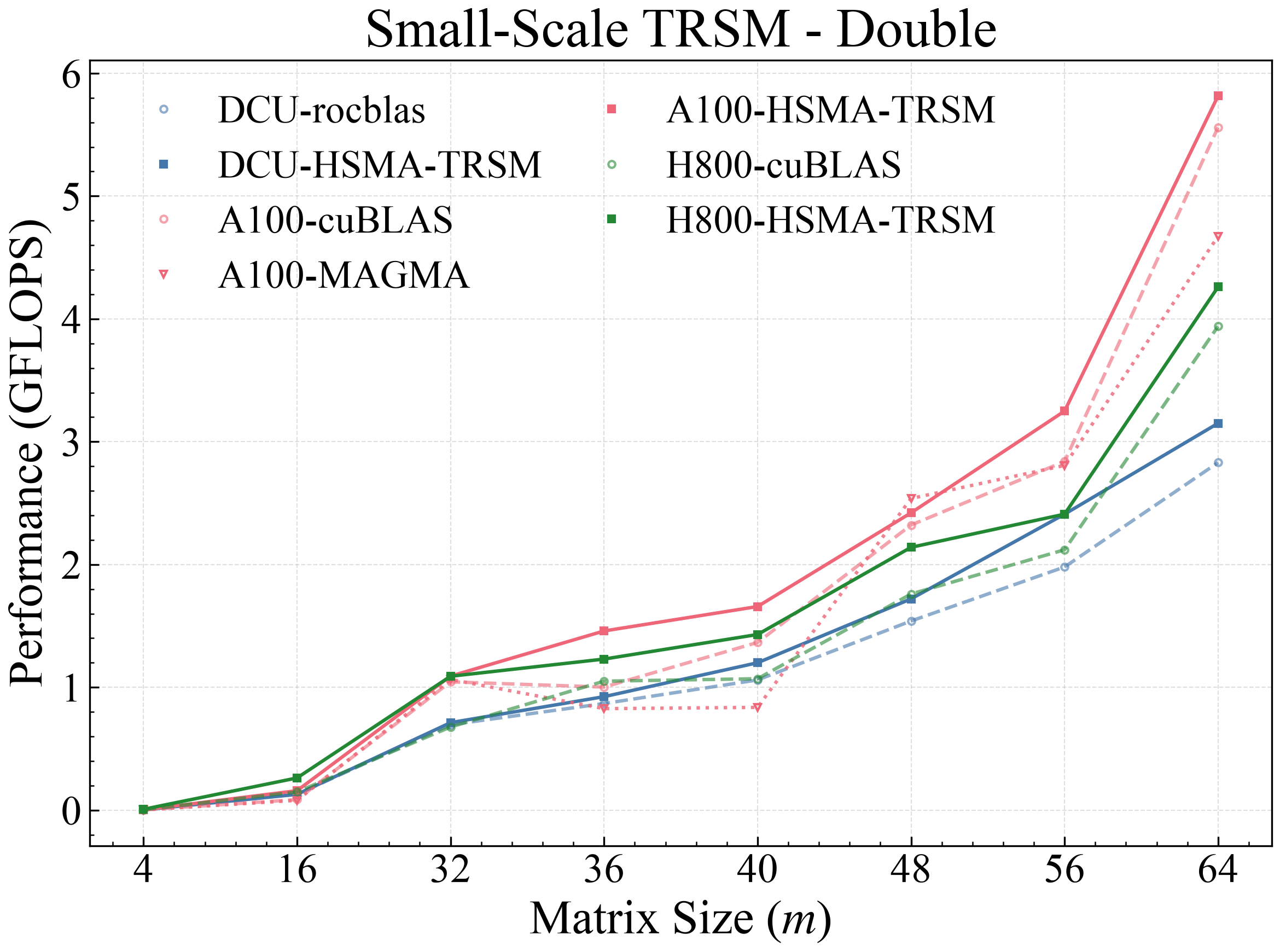}
  \caption{Double}
\end{subfigure}\\[0.2em]
\begin{subfigure}[b]{0.46\columnwidth}
  \includegraphics[width=\textwidth]{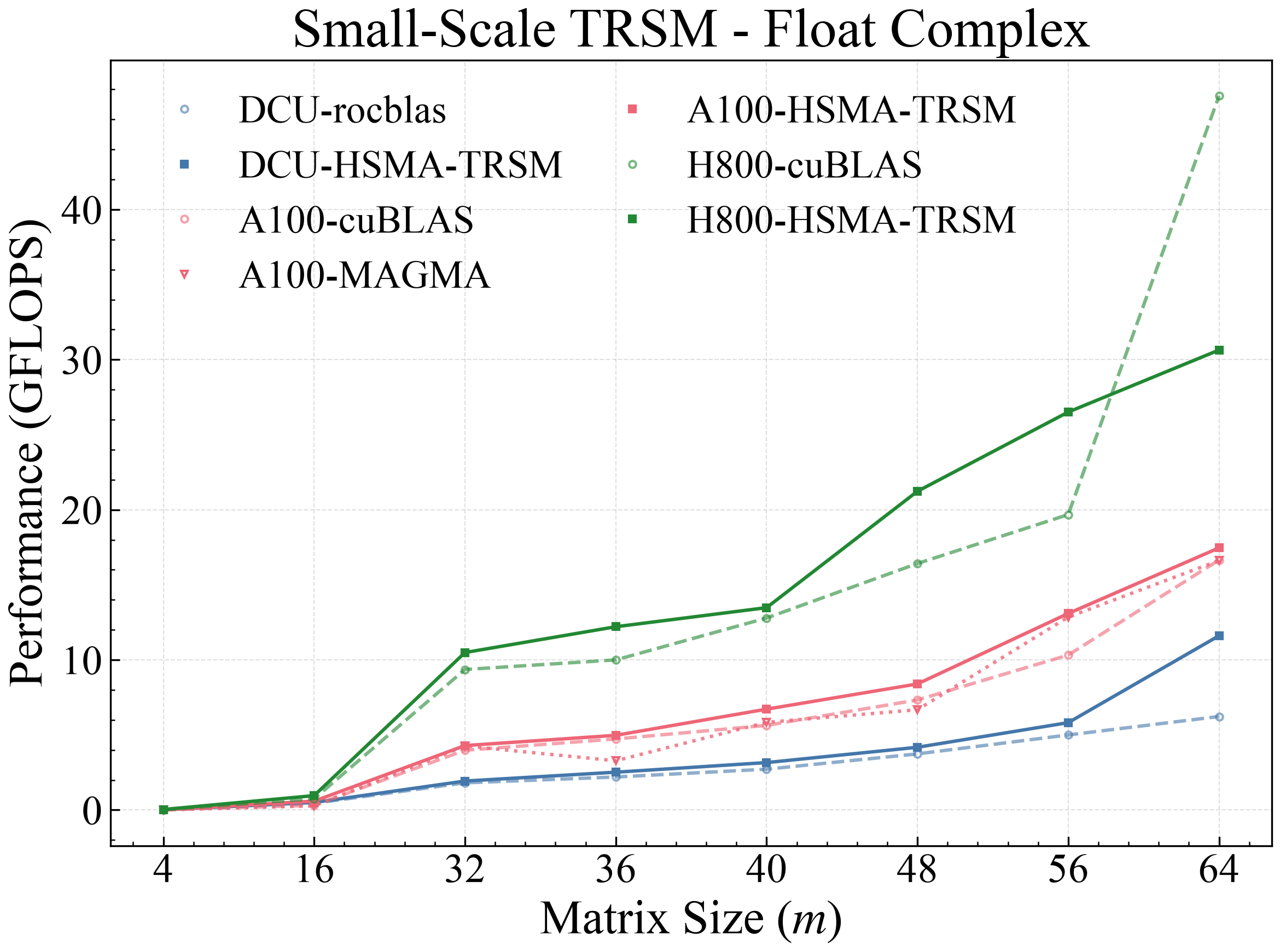}
  \caption{Float complex}
\end{subfigure}
\hfill
\begin{subfigure}[b]{0.46\columnwidth}
  \includegraphics[width=\textwidth]{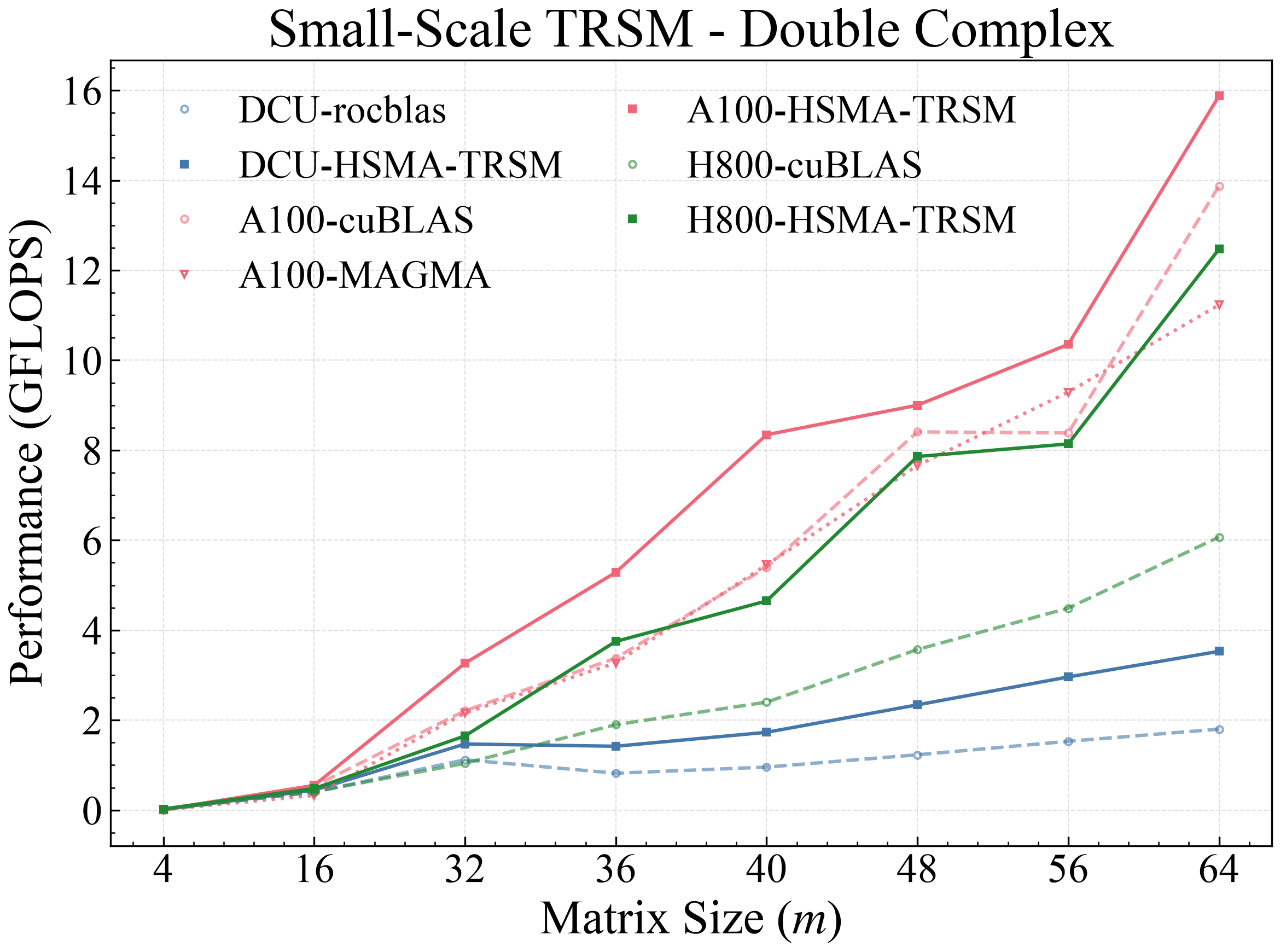}
  \caption{Double complex}
\end{subfigure}
\caption{Small-scale TRSM performance across four data types. Markers indicate measured points, and straight segments are visual guides.}
\Description{Performance charts showing GFLOPS comparison between HSMA-TRSM and vendor libraries (cuBLAS/rocBLAS) for the small-scale regime (m and n no greater than 64) across three platforms (DCU, A100, H800) and four data types (float, double, float complex, double complex). The optimized implementation shows significant improvements, especially for double complex type.}
\label{fig:small-perf}
\end{figure}

Figure~\ref{fig:small-perf} compares performance in the small-scale regime for four data types across the three platforms. The results improve most tested points, with the clearest gains near the upper end of the small-scale range and for complex types. At $m = 64$, float reaches 4.64 GFLOPS on DCU and 5.75 GFLOPS on A100, corresponding to 18.4\% and 29.2\% improvements over rocBLAS and cuBLAS, while H800 delivers the highest absolute float throughput but only a modest relative gain because the vendor kernel is already strong.

Complex types benefit more from the proposed kernels because their wider operands increase shared-memory pressure. For float complex, HSMA-TRSM reaches 11.61 GFLOPS on DCU at $m = 64$, improving over rocBLAS by 86.3\%. For double complex, the partitioned shared-memory pipeline is essential: at $m = 64$, our implementation reaches 3.53 GFLOPS on DCU, 15.88 GFLOPS on A100, and 12.47 GFLOPS on H800, corresponding to improvements of 96.1\%, 14.5\%, and 105.4\% over vendor libraries.

\begin{figure}[t]
\centering
\includegraphics[width=0.92\columnwidth]{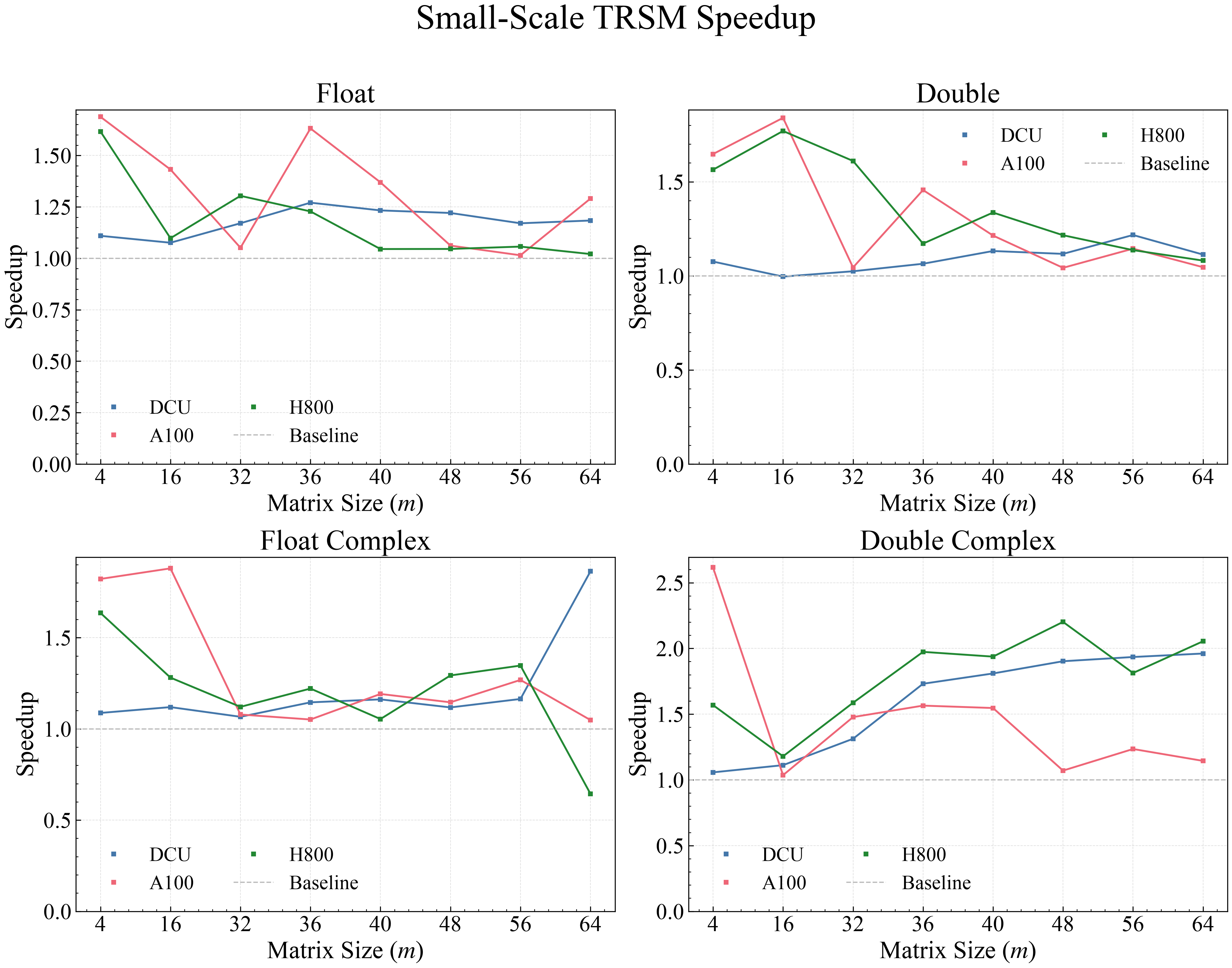}
\caption{Small-scale TRSM speedup across four data types. Markers indicate measured points, and straight segments are visual guides.}
\Description{Speedup ratios in the small-scale regime (m and n no greater than 64) with four data types (float, double, float complex, double complex) on three platforms compared to vendor libraries.}
\label{fig:small-speedup}
\end{figure}

Figure~\ref{fig:small-speedup} confirms the same trend from the speedup perspective. The largest gains appear for double complex, where the seven-stage pipeline reaches a peak speedup of 2.05$\times$ on H800 and 1.96$\times$ on DCU, while float and double mostly remain in the 1.1$\times$--1.3$\times$ range. Float complex also shows clear gains on DCU and A100, with the advantage on H800 narrowing at $m = 64$.

Compared with MAGMA on A100, HSMA-TRSM is most competitive at the upper end of the small-scale range. At $m = 64$, our implementation improves over MAGMA by 28.9\% for float, 24.6\% for double, 5.1\% for float complex, and 41.4\% for double complex. The strongest gap again appears in double complex, where the shared-memory partitioning strategy is specifically designed for the case in which a $64 \times 64$ tile of matrix $A$ already exhausts the 64KB shared-memory budget.

\subsection{Large-Scale TRSM Performance}
For problems beyond the small-scale regime, TRSM employs blocking algorithms, decomposing the problem into two phases: diagonal block inversion and GEMM update. This section evaluates the effectiveness of diagonal block optimization and adaptive blocking strategies.

\subsubsection{Impact of Diagonal Block Optimization}
In traditional implementations, diagonal block inversion requires $O(I_B^2)$ shared memory, severely limiting block size. Through decoupling optimization and double buffering mechanisms, we reduce shared memory requirements to $O(I_B)$, significantly extending the feasible block size.

\begin{table}[t]
\centering
\caption{A100 double-complex phase breakdown with block size 256}
\label{tab:phase-breakdown}
\scriptsize
\resizebox{\columnwidth}{!}{%
\begin{tabular}{ccccc}
\toprule
$m=n$ & Diag. inv. & GEMM update & Solve GEMM & Copy+other \\
\midrule
2048 & 20.7\% & 54.8\% & 20.3\% & 4.3\% \\
4096 & 8.3\% & 76.3\% & 12.9\% & 2.5\% \\
8192 & 3.7\% & 88.0\% & 7.0\% & 1.3\% \\
\bottomrule
\end{tabular}}
\end{table}

Using the float and double subfigures in Figure~\ref{fig:large-perf} as representative cases, we observe that the adaptive algorithm increases the block size as the matrix grows, e.g., from 256 at $m = 1024$ to 512 at $m = 8192$. Relative to rocBLAS's fixed-128 blocking baseline on DCU, this yields performance improvements of 39.3\% at $m = 1024$ and 23.8\% at $m = 8192$. The decoupling is not a free optimization; instead, it trades a lighter diagonal-block path for larger blocked updates. Table~\ref{tab:phase-breakdown} shows that, for A100 double complex, diagonal inversion falls from 20.7\% to 3.7\% of runtime as the problem grows, while GEMM update rises from 54.8\% to 88.0\%. This confirms that the benefit is strongest once large problems amortize diagonal preprocessing and become GEMM dominated.

\subsubsection{Overall Performance Analysis}
\begin{figure}[t]
\centering
\begin{subfigure}[b]{0.46\columnwidth}
  \includegraphics[width=\textwidth]{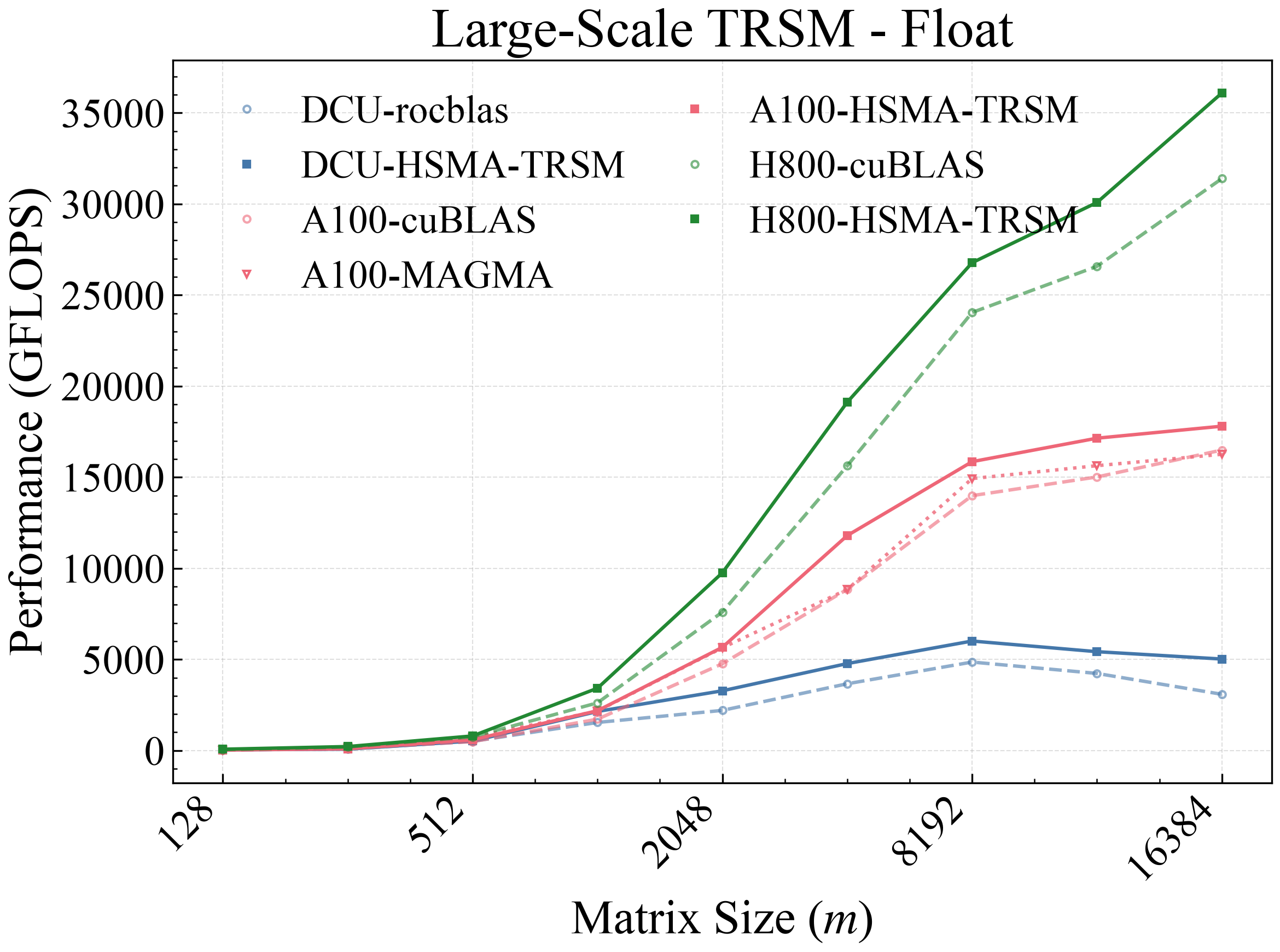}
  \caption{Float}
  \label{fig:large-float}
\end{subfigure}
\hfill
\begin{subfigure}[b]{0.46\columnwidth}
  \includegraphics[width=\textwidth]{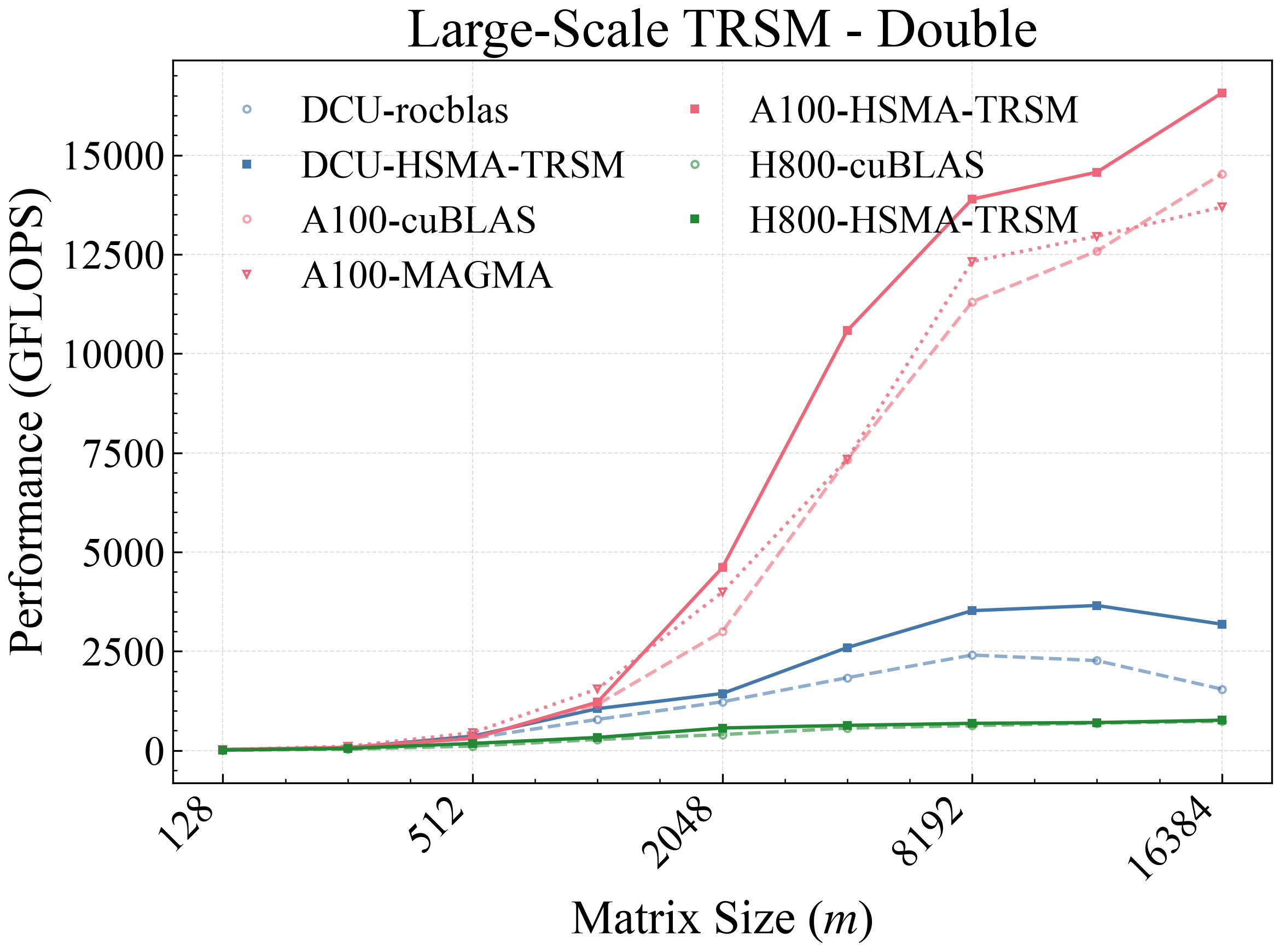}
  \caption{Double}
  \label{fig:large-double}
\end{subfigure}\\[0.2em]
\begin{subfigure}[b]{0.46\columnwidth}
  \includegraphics[width=\textwidth]{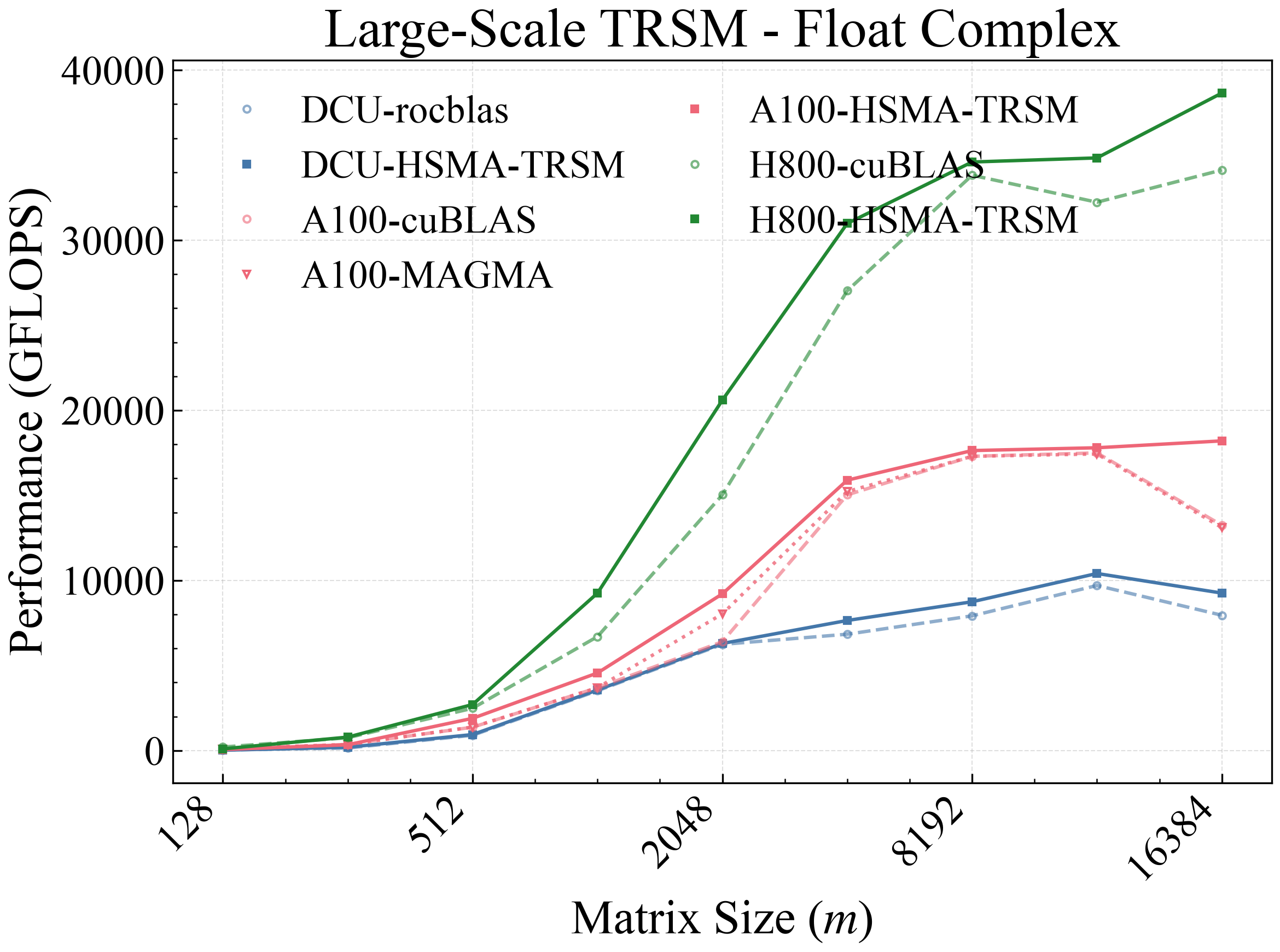}
  \caption{Float complex}
  \label{fig:large-float-complex}
\end{subfigure}
\hfill
\begin{subfigure}[b]{0.46\columnwidth}
  \includegraphics[width=\textwidth]{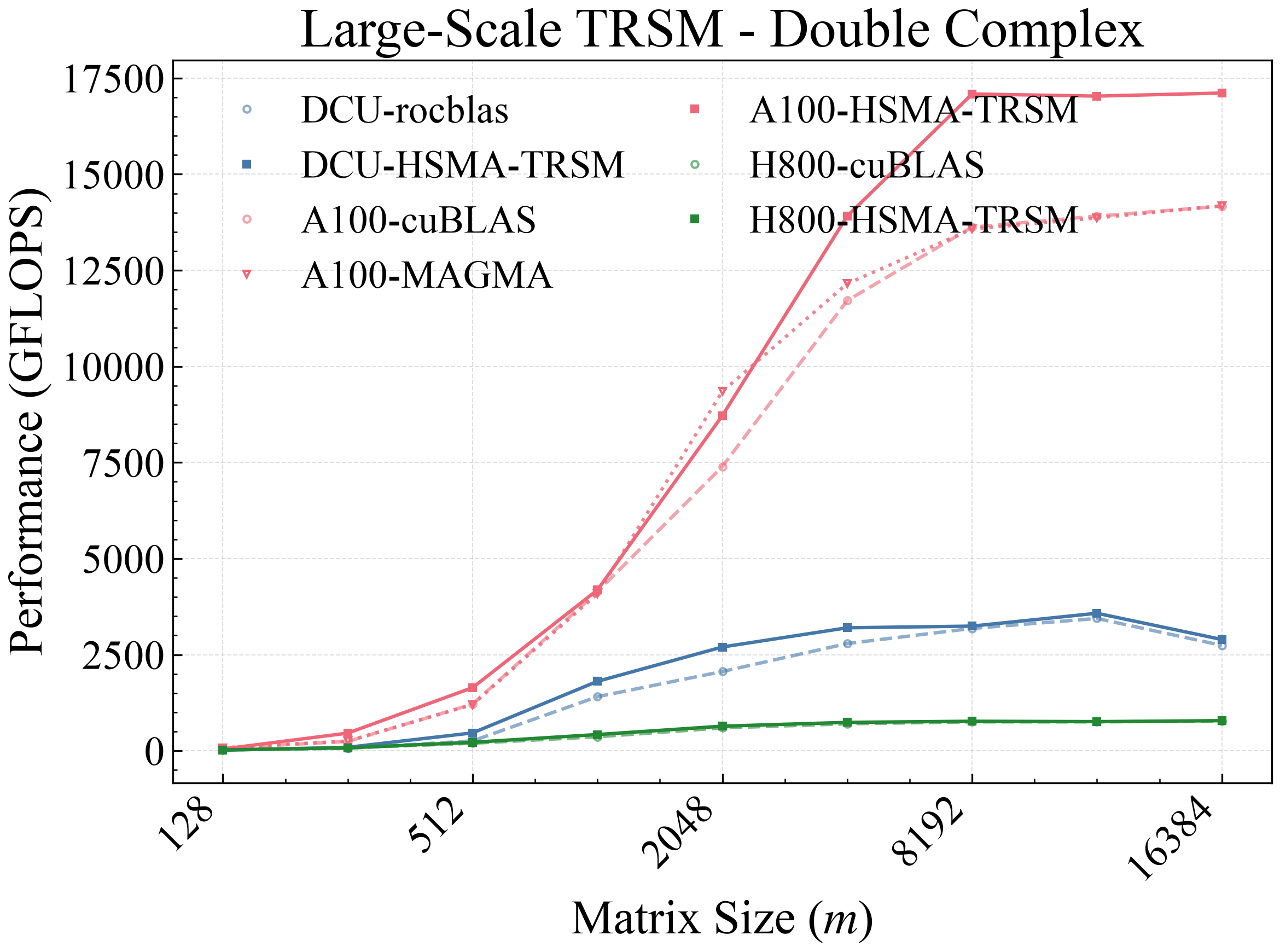}
  \caption{Double complex}
  \label{fig:large-double-complex}
\end{subfigure}
\caption{Large-scale TRSM performance across four data types. Markers indicate measured points, and straight segments are visual guides.}
\Description{A four-panel figure showing large-scale TRSM performance comparison across DCU, A100, and H800 platforms for float, double, float complex, and double complex data types. The A100 curves also include MAGMA as an additional baseline.}
\label{fig:large-perf}
\end{figure}

Figure~\ref{fig:large-perf} shows that HSMA-TRSM outperforms vendor libraries on most large-scale cases once blocked GEMM becomes dominant. On DCU, peak performance reaches 6010 GFLOPS for float and 3654 GFLOPS for double, and the speedups at $m = 16384$ remain 1.63$\times$ and 2.06$\times$, respectively. On A100, peak performance reaches 17802 GFLOPS (float), 16564 GFLOPS (double), 18205 GFLOPS (float complex), and 17111 GFLOPS (double complex), exceeding both cuBLAS and MAGMA at the upper end of the scale range. H800 delivers the highest absolute throughput for float and float complex, up to 36088 GFLOPS and 38654 GFLOPS, whereas its double-complex gains remain close to parity with cuBLAS, indicating more limited optimization headroom for that case.

\subsubsection{MAGMA Comparison}
We compare with MAGMA~\cite{tomov2010magma,magma2024} on A100. MAGMA's fixed $N_B=128$ is competitive at small and medium scales, but our adaptive strategy becomes more advantageous as the matrix grows. Representative results include 11798 versus 8837 GFLOPS for float at $m = 4096$, 10577 versus 7344 GFLOPS for double at $m = 4096$, and 17111 versus 14179 GFLOPS for double complex at $m = 16384$. This confirms that no single fixed block size is optimal across the full scale range.

\begin{figure}[t]
\centering
\includegraphics[width=0.92\columnwidth]{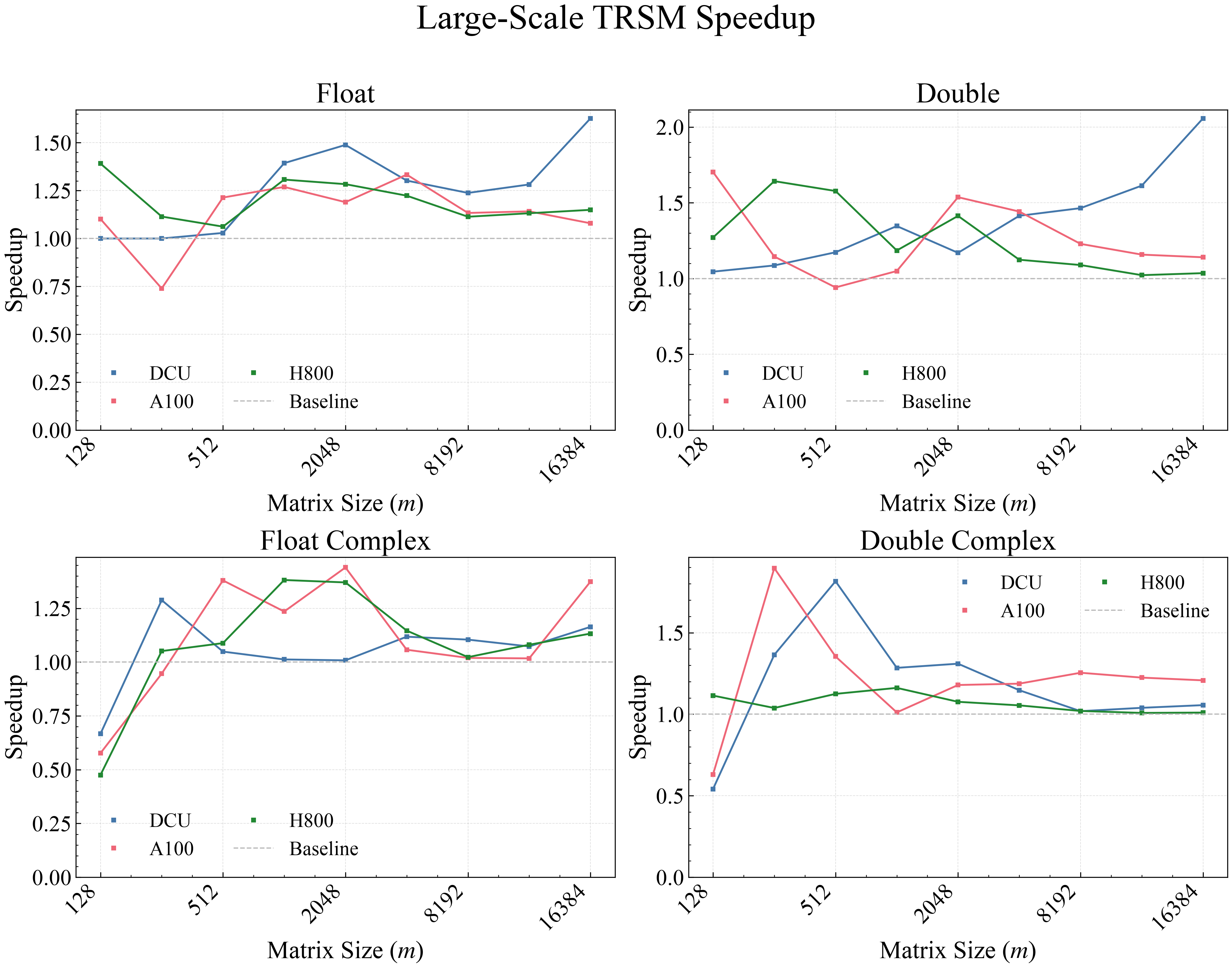}
\caption{Large-scale TRSM speedup across four data types and three platforms. Markers indicate measured points, and straight segments are visual guides.}
\Description{Speedup ratios of large-scale TRSM on DCU, A100, and H800 platforms relative to vendor libraries (cuBLAS/rocBLAS) across float, double, float complex, and double complex data types.}
\label{fig:large-speedup}
\end{figure}

Figure~\ref{fig:large-speedup} shows the same trend from the speedup perspective: DCU exhibits the largest gains for real types, A100 provides the most consistent improvements for complex types, and H800 stays close to parity for double complex.

\subsection{Analysis and Discussion}
Cross-platform performance differences stem from both hardware throughput and the effective per-block on-chip storage budget used by our kernels. In our target configuration, DCU provides a 64KB per-block LDS budget, whereas A100 and H800 use a 48KB per-block shared-memory budget. This larger per-block budget favors fine-grained on-chip staging on DCU, while A100 and H800 provide higher absolute compute throughput and memory bandwidth, making larger blocked updates more effective once the computation becomes GEMM-dominated. This behavior is consistent with the roofline analysis in Figure~\ref{fig:roofline}: small-scale TRSM has low arithmetic intensity and is sensitive to shared-memory reuse and memory-computation overlap, whereas large-scale blocked TRSM increasingly spends time in GEMM-like updates. Therefore, strong vendor-library cases indicate limited remaining headroom rather than a failure of the proposed optimizations. Our adaptive framework targets the regimes where shared-memory pressure and fixed blocking choices still leave exploitable room.

\section{Conclusion and Future Work}

This paper proposes HSMA-TRSM, a hierarchical shared memory-aware optimization framework for left-side lower-triangular TRSM on NVIDIA GPUs (A100, H800) and Hygon DCU Z100. In the small-scale regime ($m, n \leq 64$), HSMA-TRSM uses a pipelined compute-memory overlap design and a seven-stage dual thread-group pipeline for double complex, achieving a peak 2.05$\times$ speedup over vendor libraries.

For problems beyond the small-scale regime, HSMA-TRSM combines diagonal block decoupling with adaptive blocking, reducing the shared-memory footprint of diagonal block inversion from $O(I_B^2)$ to $O(I_B)$ and enabling larger blocked updates. It achieves up to 51.9\% improvement for double precision on DCU and maintains 1.63$\times$ to 2.06$\times$ speedups at $m = 16384$ for real types, while the gains for mature vendor kernels and some complex large-scale cases are more modest. Future work includes kernel fusion for very small scales, batched TRSM, and extensions to sparse and mixed-precision settings.

\begin{acks}
This work was supported by the
\grantsponsor{GS501100012166}{National Key Research and Development Program of China}{https://doi.org/10.13039/501100012166}
under Grant No.~\grantnum{GS501100012166}{2023YFB3001700}.
\end{acks}

\bibliographystyle{ACM-Reference-Format}
\bibliography{references}

\end{document}